\documentstyle [prl,aps,multicol,amstex,epsfig]{revtex}
\begin{document}
\draft
\hyphenation{following}

\title{ Symmetry Analysis of Second Harmonic Generation at Surfaces of
Antiferromagnets }
\author{M. Trzeciecki$^{1,3}$, A. D\"ahn $^2$, and W. H\"ubner$^1$}
\address{$^1$Max-Planck-Institut f\"ur Mikrostrukturphysik, Weinberg 2,
D-06120 Halle, Germany}
\address{$^2$Institute for Theoretical Physics, Freie Universit\"at 
Berlin, Arnimallee 14, D-14195 Berlin, Germany}
\address{$^3$Institute of Physics, Warsaw University of Technology, Koszykowa 
75, 00-662 Warsaw, Poland}
\date{\today}
\maketitle
\begin{abstract}
Using group theory we classify the nonlinear magneto-optical response at 
low-index surfaces of fcc antiferromagnets, such as NiO. Structures consisting 
of one atomic layer are discussed in detail. We find that optical second 
harmonic generation is sensitive to surface antiferromagnetism in many cases.
We discuss the influence of a second type of magnetic atoms,
and also of a possible oxygen sublattice distortion on the output signal.
Finally, our symmetry analysis yields the possibility of antiferromagnetic 
surface domain imaging even in the presence of magnetic unit-cell doubling.
\end{abstract}
\pacs{78.20.Ls, 75.30.Pd, 75.50.Ee, 42.65.-k}
\vskip 0.5in
\begin{multicols} {2}
\section{Introduction}
\noindent Optical Second Harmonic Generation (SHG) has been proven to be a very 
useful technique 
for the investigation of ferromagnetism at surfaces. The obvious question is if 
this technique can also yield some new information in the case of more general
spin configurations, such as antiferromagnetic (AF) ordering. An 
experimental answer to this question has been 
provided by Fiebig {\em et al.} \cite {ref18}, who obtained a pronounced 
optical contrast from AF 180$^\circ$ domains of rhombohedral bulk Cr$_2$O$_3$.
The authors attributed this contrast to the interference of magnetic and 
electric dipole contributions, the latter being present only below the N\'eel 
temperature. Since it is known that, in {\em cubic} materials, within the 
electric dipole approximation, optical SHG originates only from surfaces, 
interfaces, or thin films, an important 
question is if SHG is also sensitive to antiferromagnetism at surfaces of 
cubic antiferromagnets. In this paper, we will show that the surface of a cubic 
material can lower the symmetry of an AF fcc crystal (two-sublattice 
antiferromagnet) in a way similar to
the trigonal distortion in a four sublattice antiferromagnet Cr$_2$O$_3$. 
Besides, even the imaging of AF {\em domains} is possible also for many cubic 
materials that exhibit unit-cell doubling. 

The first theoretical explanation of {\em linear} magneto-optic effects in 
ferromagnets has been given by Argyres \cite{ref25} in the 50s. He used linear 
response theory for current-current correlation functions. His 
microscopic explanation was already based on the combination of spin-orbit
and exchange coupling. Experimental techniques for the detection of AF domain 
{\em walls} using linear optics in some special geometries were elaborated a few 
years later~\cite {ref7zref27}. The {\em interior} of the domains has been 
visualized in piezoelectric AF crystals using a linear magneto-optical effect 
\cite {ref4zref17}. However, linear optical experiments 
suffer from mixing the desired signal with a contribution from other linear 
effects, such as birefringence or dichroism. A review of linear optical 
experimental methods for the investigation of AF domains is given by 
Dillon~\cite{ref17}. 

The observation of domain structure in antiferromagnets is more complicated 
than in ferromagnetic materials since the reduction of the spatial symmetry is, 
unlike for ferromagnets, not linked to an imbalance in the occupation of 
majority and minority spin states. On the basis of group theoretical 
considerations, Brown {\em et al.} \cite {ref28} proposed  the use of linear 
optical effects, namely gyrotropic birefringence, for the observation of AF 
domains 
related to each other by the space-inversion operation. A theoretical review of 
effects found by a group-theoretical approach is presented by Eremenko and 
Kharchenko \cite {ref13}. They performed a comprehensive study of linear optical 
effects for various AF materials. Another effect proposed recently by 
Dzyaloshinskii {\em et. al.} \cite{ref41} gives a possibility to 
detect antiferromagnetism taking advantage from optical path differences from 
antiferromagnetically coupled but intrinsically ferromagnetic planes.

{\em Nonlinear} optics exhibits an additional degree of freedom, since its 
elementary process involves three photons instead of two in linear optics. For 
that reason, some authors, e.g. Fr\"ohlich \cite{ref40} suggested the 
application of nonlinear optics even for k-selective spectroscopy, since 
multi-photon phenomena allow for the ``scanning'' of a small part of the 
Brillouin zone, at least for 
semiconductors. Recently, non-linear optics has attracted more and more 
attention for the investigation of magnetism due to its enhanced sensitivity to 
twodimensional {\em ferromagnetism} \cite {ref53}. The magnetic effects are 
usually much stronger than in linear optics (rotations up to 90$^\circ$, 
pronounced spin polarized quantum well state oscillations \cite{ref64,ref65}, 
magnetic contrasts 
close to 100$\%$) \cite {ref22,pustu}. An example of ferromagnetic effects 
measurable only by SHG deals with the existence of surface magnetism in very 
thin films of Fe/Cu(001) and is given in Ref. \cite {ref31}. Nonlinear optical 
effects were invoked to explain the behavior of lasers in magnetic fields 
\cite{ref55}, to investigate high 
temperature superconductors \cite{ref67,ref30}, and to study structures composed 
from alternately ferro- and antiferromagnetically ordered thin films 
\cite{ref34}. One of the first theoretical investigations
of the possibility to apply nonlinear optics to {\em antiferromagnetism} 
was performed by Kielich and Zawodny \cite{Kieliszek}. However, the first 
experiments concerning the detection of the AF domains in materials such as 
Cr$_2$O$_3$ were carried out only recently~\cite {ref4,ref12}. Already in the 
70s, it has been proposed \cite{ref10} that experimental 
studies of dc magnetic and electric field-induced SHG could become 
an effective method of determining the crystal structure of solids, the 
symmetry of which cannot be investigated by other methods. Extending this idea 
towards surface crystallography provides us with a new technique for  
determining the spin configuration in a given surface structure. In turn, it
permits to use a known magnetic configuration for the determination of the 
surface structure. All the mentioned effects are more difficult or even 
impossible to obtain in linear optics, and moreover other linear methods like 
neutron scattering have difficulties to probe AF spin configurations.
 
The nonlinear magneto-optical susceptibility tensor $\chi^{(2\omega)}_{el}$ (the 
source for SHG within the electric dipole approximation) has predominantly been 
investigated from the symmetry point of view. A classification following this 
approach, with tensors of a rank up to six, has been performed by Lyubchanskii 
{\em et al.} \cite {ref19,ref20,ref21,ref22,ref23}. 
In Ref. \cite {ref22} the authors include the magnetization-gradient terms and 
apply the group-theoretical classification to higher-rank susceptibility 
tensors. This approach then allows them to study the thickness and the character 
(Bloch vs. N\'eel type) of domain walls. An attempt by Muthukumar {\em et. al.} 
\cite {ref27} to calculate the $\chi^{(2\omega)}_{el}$ tensor elements for the 
antiferromagnetic Cr$_2$O$_3$ both from group theory as well as {\em from the 
microscopic point of view} 
is rather unique. They implemented a (CrO$_6$)$_2$ cluster, thus taking into 
account only half of the spins present in the elementary magnetic cell. In this 
approximation they explained the SHG from Cr$_2$O$_3$ as observed by Fiebig {\em 
et al.} \cite{ref18} and they were able to give a quantitative estimate for 
that. Tanabe {\em et al.} \cite {ref56}, however, pointed out that the 
occurrence of purely real or imaginary values of the tensor elements plays a 
decisive role for the existence of SHG from this substance.
They found that for a (CrO$_6$)$_2$ cluster SHG can take place only in the case
where the tensor elements are imaginary, and thus should vanish in Muthukumar's
approximation. They proposed to take into account the full unit cell with four 
inequivalent Cr ions including their ``twisting'' interaction with the 
environment. However Tanabe {\em et al.} neglected the dissipation in the 
process of SHG \cite {RemTanabe}, which is a rather crude approximation. In 
general, taking into account the dissipation makes the $\chi^{(2\omega)}_{el}$ 
tensor elements complex  and invalidates their separation in purely real and 
imaginary ones \cite{unpublished}.

Lifting the inversion symmetry of a crystal is the source for SHG. Lyubchanskii 
{\em et al.} \cite {ref19,ref21} suggested crystal lattice deformations and 
displacements as possible reasons for SHG from YIG films. In the case of 
Cr$_2$O$_3$ and YBa$_2$Cu$_3$O$_{6+\delta}$, described by Lyubchanskii
{\em et al.} \cite {ref20,ref21}, AF ordering lowers the symmetry of an 
otherwise centrosymmetric crystal. In this paper, however, we rely on the idea 
that, rather than 
lowering the crystal symmetry in the bulk, SHG may also result from the 
breaking of inversion symmetry at the surface of a bulk inversion-symmetric 
system. 

Magnetically active oxide layers are of importance for the construction of TMR 
(tunneling magnetoresistance) devices, where a trilayer structure is commonly 
used. The central layer 
of TMR devices consists of an oxide sandwiched between a soft and a hard 
magnetic layer (these two layers are often composed from the same material but 
of different thicknesses). For these technological applications it is necessary 
to develop a technique to study buried oxide interfaces. Such a technique can be 
SHG. One of the most promising materials for the mentioned devices is NiO. 
However, to the best of our knowledge, the understanding of its detailed spin 
structure is scarce - even the spin orientation on the ferromagnetically ordered 
(111) surfaces is not known. The technique presented here can shed some light on 
that issue.

Our paper is organized as follows: in Sec. II we 
present our methods for obtaining sets of nonvanishing 
$\chi^{(2\omega)}_{el}$ tensor elements. In Sec. III we present the results
of our analysis, first for the nondistorted surface of a simple fcc structure 
(subsection III.A), then for the the distorted one (III.B). Subsequently, we 
discuss the influence of a second kind of magnetic atoms (III.C) and of oxygen 
sublattice distortion (III.D). The issue of domain imaging is addressed in 
subsection III.E. Possible experimental geometries allowing for the detection of 
the mentioned structures and effects are discussed in Sec. IV. The conclusions 
are presented in Sec. V.

\section{Theory}
\noindent Based on group theory, D\"ahn {\em et al.}~\cite{dahn} 
proposed a new nonlinear magneto-optic Kerr effect (NOLIMOKE) at the surface of 
cubic antiferromagnets. They also gave an example of an antiferromagnetic 
structure (NiO) and an optical configuration, where this new effect could be 
observed. Here, we perform a complete group-theory based analysis of collinear 
AF fcc low-index crystal surfaces. Surfaces of other crystal structures are as 
well described by our theory provided they are similar to
fcc crystal surfaces, i.e. squares or hexagons. The results can be used to 
detect the magnetic order of a specific surface under investigation and allow 
for the determination of the surface spin configuration in some important cases. 
However, in order to calculate the SHG yield quantitatively, it is necessary to 
go beyond the present study and use electronic calculations of the nonlinear 
susceptibility. Group theory can give a unified picture of different 
experimental observations and predict new effects \cite {symmetry}, while the
microscopic origins of the observed phenomena may remain unclear. 
In order to be clear with respect to the essential notion of time reversal 
we would like to emphasize the point of view taken in this paper in the 
beginning. Here, we do not divide $\chi^{(2\omega)}_{el}$ into even and odd 
parts in the magnetic order parameter. Instead, the behavior of 
$\chi^{(2\omega)}_{el}$ with respect to the magnetic order parameter 
(which for ferromagnetic materials corresponds to the dependence of 
$\chi^{(2\omega)}_{el}$ on magnetization) is fully taken into account by the 
considerations of the magnetic point group. At no stage of our consideration we 
invoke the notion of time reversal, consequently we do not apply the 
characterization of the susceptibility $\chi^{(2\omega)}$ as c-tensor (changing 
its sign in the time-reversal operation) or i-tensor (invariant under the 
time-reversal operation) \cite{unpublished}.

Before we start our group theoretical classification of the nonlinear optical
susceptibilities of AF surfaces we would like to emphasize the following four
important points:\\
(i) We are not interested in effects resulting from the {\em optical path 
difference} from adjacent crystal planes which are ferromagnetically ordered but 
only antiferromagnetically coupled to each other. We do not consider this as an 
intrinsic AF effect. \\
(ii) Cubic crystals that we are interested in reveal a center of inversion in 
the para-, ferro-, and all antiferromagnetic phases. Thus, within the electric 
dipole approximation, the SHG signal from the bulk vanishes.\\
(iii) While in principle linear optical methods can be sensitive to the 
presence of a spin structure, in practice they are not useful because, within 
the group theoretical approach, they 
cannot distinguish the AF phase from either paramagnetic or ferromagnetic, nor 
can they distinguish different AF configurations from each other. They have to 
resort to methods like lineshape analysis, where no strong statements 
characteristic for symmetry analysis can be made. \\
(iv) Although the tensor elements for all the magnetic point groups are known 
and tabulated in the literature (e.g. \cite{birss}), the connection between the 
different spin configurations described by us and the mentioned symmetry groups 
has not been made, except for some easy cases \cite{dahn}. Thus, for SHG from 
antiferromagnetic surfaces there has been up to now no connection between the 
group theoretical classification and the real situations found in experiments. 

The following part of the text should explain the fundamentals of applying 
NOLIMOKE observations to investigate antiferromagnetism of surfaces. 

Now we turn to SHG, the source of which is the nonlinear electrical polarization 
$P^{(2\omega)}_{el}$ given by:
\begin{equation}
P^{(2\omega)}_{el}=\epsilon_0 \chi^{(2\omega)}_{el}:E^{(\omega)}E^{(\omega)} .
\end{equation}
Here, $E^{(\omega)}$ is the electric field of the incident light, while 
$\chi^{(2\omega)}_{el}$ denotes the nonlinear susceptibility within the electric 
dipole approximation, and $\epsilon_0$ is the vacuum permittivity.
The intensity of the outgoing SHG light is \cite{Sipe}:
\begin {multline}
\label{eqGlown}
I^{(2\omega)}\sim (I_0)^2
\Bigl [ F(\Theta ,\Phi , 2\omega) \times \\
\times \begin{pmatrix}
\chi_{xxx}&\chi_{xyy}&\chi_{xzz}&\chi_{xyz}&\chi_{xzx}&\chi_{xxy}\\
\chi_{yxx}&\chi_{yyy}&\chi_{yzz}&\chi_{yyz}&\chi_{yzx}&\chi_{yxy}\\
\chi_{zxx}&\chi_{zyy}&\chi_{zzz}&\chi_{zyz}&\chi_{zzx}&\chi_{zxy} \end{pmatrix}
\times \\ \times
f(\vartheta ,\varphi ,\omega) \Bigr ]^2
\end{multline}
where $I_0$ is the intensity of the incident light, $F()$ ($f()$) describe 
Fresnel and geometrical factors for the incident (reflected) light, $\vartheta$ 
and $\Theta$ angles of incidence and reflection, respectively 
($\vartheta$=$\Theta$), and $\Phi$ ($\varphi$) is output (input) polarization 
angle. According to Neumann's principle, ``any type of symmetry which is 
exhibited by the crystal is possessed by every physical property of the 
crystal'' \cite{birss}. To examine these physical properties, we determine the 
magnetic point group of the crystal lattice, thus determine its symmetries. The 
same symmetries must leave the investigated property tensor (in our case the 
nonlinear electric susceptibility~$\chi^{(2\omega)}_{el}$) invariant. This fact 
is 
mathematically expressed by the following condition:
\begin{equation}
\label {eq2}
\chi^{(2\omega)}_{el,i'j'k'}=l_{i'i}l_{j'j}l_{k'k} 
\chi^{(2\omega)}_{el,ijk},\,\,
\,\,
  i,j,k,i',j',k'=x,y,z .
\end{equation}
Here, $\mathit{l_{n,n'}} (n = i,j,k, n'= i',j',k',)$ is a representation of an
element of the magnetic point group describing the crystal. For symmetry 
operations including the time reversal there should be an additional ``$\pm$'' 
sign 
in Eq.(\ref {eq2}), but we do not use it here since we exclude the time reversal 
from our consideration. In particular, from 
Eq.(\ref {eq2}) it follows immediately that polar tensors 
of odd rank (such as~$\chi^{(2\omega)}_{el}$ )
vanish in inversion symmetric structures. This explains why SHG is possible 
only at surfaces and interfaces, where this symmetry is broken.
For a given spin configuration we apply Eq. (\ref {eq2}) for every symmetry
operation exhibited by the system. Thus, each of these symmetries gives rise
to a set of 27 equations with 27 unknown elements of the tensor 
$\chi^{(2\omega)}_{el}$. This set can be reduced to 18 equations, since
\begin {equation}
\label {eq3}
\chi^{(2\omega)}_{el,ijk}=\chi^{(2\omega)}_{el,ikj},
\end{equation}
which expresses the equivalence of the incident photons of frequency $\omega$, 
see also the reduced notation in Eq. (\ref{eqGlown}). The analytic solution of 
even this reduced set of equations seems cumbersome, but the set can be split 
into several decoupled subsets. For example, an obvious subset in every case is 
the equation $\chi_{zzz} = \chi_{zzz}$, this tensor element occurs nowhere else. 
The rank of other subsets is, for our cases, never higher than six. In this 
manner, one may obtain a set of forbidden elements of the susceptibility tensor 
as well as relations between existing ones.

\section{Results}

First, we will define the notions of ``phase'', ``case'', and ``configuration'', 
used henceforth to classify our results. ``Phase'' describes the magnetic phase 
of the material, i.e. paramagnetic, ferromagnetic, or AF. Secondly, the 
word ``configuration'' is reserved for the description of the magnetic ordering 
of the surface. It describes various possibilities of the spin ordering, which 
are different in the sense of topology. We describe up to 18 AF 
configurations, denoted by little letters a) to r), as well as several 
ferromagnetic configurations, denoted as ``ferro1'', ``ferro2'', etc. The number 
of possible configurations varies depending on surface orientation. Thirdly, we 
describe different ``cases'', i.e. additional structural 
features superimposed on the symmetry analysis. ``Case A'' does not have such 
additional features. In ``case B'' we address distortions of the lattice. ``Case 
C'' deals with two kinds of magnetic atoms in an undistorted lattice. In ``case 
D'' we take into account a distorted sublattice of nonmagnetic atoms, keeping 
the magnetic sublattice undistorted. All the analysis concerns collinear 
antiferromagnets, with one easy axis.

The tables show the SHG response types for each configuration. The various 
response types are encoded by a ``key'', which is then decoded in Tab. 
\ref{keytab}. This table presents the symmetries, domain operations, and 
nonvanishing tensor elements for each response type. This is done in order to 
shorten the overall length of tables, because a given response type can appear 
in several different cases.

\begin{figure}
\epsfig{file=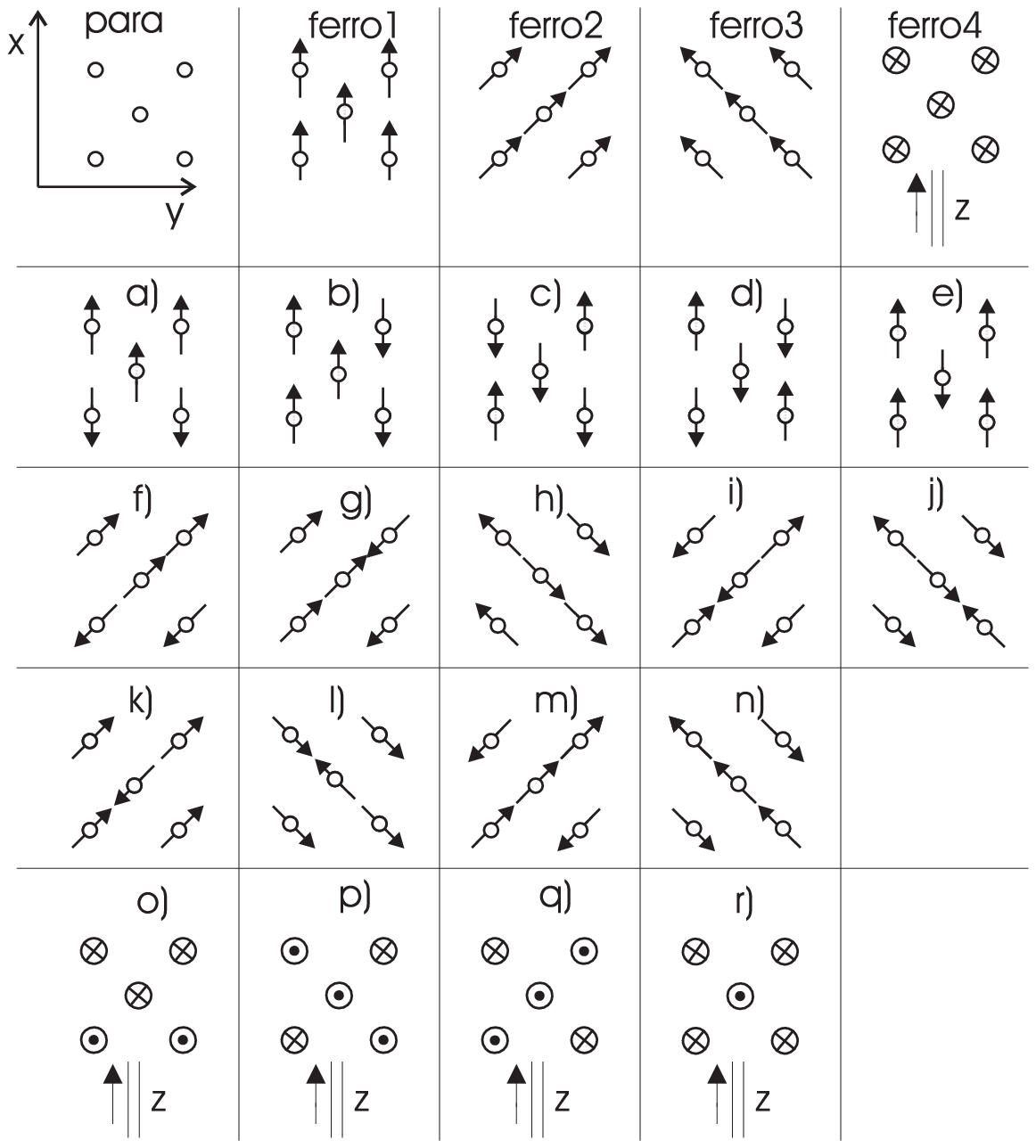}
\caption[fig1]{\label{fig1}Spin configurations of an fcc (001) surface. Except 
for confs. ``ferro4" and  o) - r), the arrows always indicate in-plane 
directions of the spins. In confs. ``ferro4" and o) - r) $\odot$ ($\otimes$) 
denote spins pointing along the positive (negative) z-direction, respectively.}
\end{figure}

Several spin structures depicted in Fig. \ref{fig1} and Fig. \ref{fig3} are 
distinct configurations only in case B, and they are addressed in the tables 
that concern only this case. For the rest of the cases they are domains of 
other, fully described configurations, thus they are left out in these cases. 
The philosophy of the paper is that, to save some space, we show the spin 
structure in one figure for each surface (Fig. \ref{fig1}, \ref{fig2}, and 
\ref{fig3}) for all the four cases (A-D), and depict the effects taken into 
account in the cases B-D only for the paramagnetic phase (Fig \ref{fdis111}, 
\ref{fnierow}, and \ref{ftlen}). Table \ref{keytab} also contains the 
information on the parity of the nonvanishing tensor elements: the odd ones are 
printed in 
boldface. In some situations an even tensor element (shown in lightface) is 
equal to an odd element (shown in boldface), this means that this pair of tensor 
elements is equal in the domain which is depicted on the corresponding figure, 
but they are of opposite sign in the other domain. This happens in the 
structures where two pairs of domains are possible (two distinct entries in 
Table \ref{keytab}). The tensor elements that change their parity in the domain 
operation which is the inverse of the displayed one are shown in italic font. 
For example, the entry j) of Table \ref{keytab} shows a tensor element {\it 
xxx}, which is even under the operation $4_z$, this means that this tensor 
element is odd under $-4_z$. This strange at the first sight behavior of tensor 
elements is caused by the fact that under these operations, tensor elements are 
not mapped on themselves. In our example, after applying $4_z$ the tensor 
element $xxx$ becomes $yyy$, without changing its sign. If we now apply $-4_z$, 
$yyy$ (which is now even) becomes $xxx$, again without changing the sign.

The parity of the elements has been checked in the operations $2_z$, $4_z$, and 
in the operation connecting 
mirror-domains to each other (for the definition of the mirror-domain structure 
see subsection E). The domain operation(s) on which the parity depends is (are), 
if applicable, also displayed in this Table. If two or more domain operations 
have the same effect, we display all of them together. To make the Table 
\ref{keytab}
shorter and more easily readable some domain operations (and the corresponding 
parity information for the tensor elements) are not displayed, namely those that 
can be created by a superposition of the displayed domain operations. We also do 
not address the parity of tensor elements in the $6_z$ nor $3_z$ operations for 
(111) surfaces nor any other operation that ``splits'' tensor elements, although 
these operations also lead to a domain structure \cite{par111}. As will be 
discussed later (subsection E) it is possible to define a parity of the tensor 
elements for the $3_z$ and $6_z$ operations, however the tensor elements then 
undergo more complicated changes. The situations where the parity of the tensor 
elements is too complicated to be displayed in the Table are indicated by a 
hyphen in the column ``domain operation''. For some configurations, none of the 
operations leads to a domain structure - in those configurations we display the 
information ``one domain''. The reader is referred to the Appendix for the 
particularities of the parity check.

As far as the first layer is concerned, we address all the spin configurations 
of the low index surfaces of fcc 
antiferromagnets, with magnetic order vector lying in plane or perpendicular to 
it and antiferromagnetic coupling between nearest neighbors. For the (001) 
surfaces we also discuss the configurations, where the antiferromagnetic 
coupling exists between the second-nearest neighbors (configurations a), b), c), 
f), and o), along with d), g), and h) for case B.). We do not consider the 
coupling to the third and further 
neighbors. This would not give rise to configurations of different symmetries in 
two dimensions. It may at most replace spins by grains (blocks) of spins in the 
configurations described by us.

Throughout this paper we take into account the spin structure only of the first 
(uppermost) atomic layer. This is sufficient to study all the symmetries of 
(001) and (110) surfaces both in the paramagnetic and ferromagnetic phases. For 
the (111) surface it is necessary to recognize the atomic positions (but not the
spins) in the second layer for the same purpose. For the sake of completeness we 
also present a study of (111) surfaces without this extension. However, in the 
antiferromagnetic phase, the spin structure of the second and deeper layers 
plays a role in determining the symmetry of the surface. This is presented in 
this paper using the (001) surface as an example. For the (110) and (111) 
surfaces it will be published elsewhere \cite{Noptipap}. These 
structures can serve as simple models for deriving predictions for more 
complicated cases, while the full consideration of the second layer would not 
bring any new interesting results. Taking into account the spin structure of the 
second layer (deeper layers do not bring up anything new to the analysis) 
results in creating several (up to two for the (001) surface and three for the 
(111) surface) configurations out of each one addressed here by us. The symmetry 
of these configurations may remain the same or be lowered (sometimes even below 
the symmetry of the ferromagnetic phase) with respect to the ``two-dimensional'' 
configurations they are generated from. Consequently the distinction of the 
configurations from each other may be limited, but the possibility to detect the 
magnetic phase is not severely affected. Also our remarks on domain imaging 
remain valid. However the number of domains is increased, thus the possibility 
to identify each of them might be hampered. 

Consequently, one can state that the symmetry of an AF surface depends on two 
atomic layers. They are also necessary (and sufficient) to define AF bulk 
domains. As will be presented in our results, SHG can probe both these layers on 
AF surfaces. 

\subsection{Equivalent atoms}

\noindent 
The predicted new nonlinear magneto-optical effects result from the 
fact that the magnetic point groups of antiferromagnetic configurations are 
different from those describing paramagnetic or ferromagnetic phases of the 
same surface. Since, depending on the magnetic phase, different tensor elements 
vanish, it is possible to detect antiferromagnetism optically by varying the 
polarization of the incoming light. 

The current subsection discusses nonvanishing elements of the nonlinear
susceptibility tensor for an fcc crystal consisting of only one kind of 
magnetic atoms. The influence of nonmagnetic atoms in the material will be 
discussed later. The configurations considered here are ``ferro1", ``ferro2", 
``ferro4", a), b), c), e), f), i), k), m), o), p), and r) for the (001) surface 
(see Fig. \ref{fig1}), ``ferro1", ferro3", ``ferro5", a), c), f), i), and k) for 
the (111) surface (see Fig. \ref{fig3}), and all configurations depicted in Fig. 
\ref{fig2} for the (110) surface. Other depicted spin structures form domains of 
these configurations and are not referred to in this subsection nor in the 
tables concerning the current subsection \cite{leftout}.

\begin{figure}
\epsfig{file=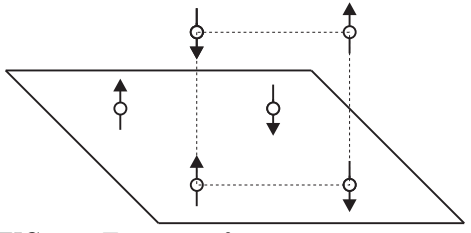}
\caption{\label{fprim} Top view of a spin structure on a (001) surface. The 
dashed line depicts a conventional unit cell, while the solid one outlines the 
primitive unit cell.}
\end{figure}

All possible configurations (confs.) of a fcc (001) surface are shown in 
Fig.~\ref{fig1}, which displays the conventional rather than magnetic unit 
cells. However, these are sufficient to fix the spin configuration of the whole 
surface imposing of the following ``convention'': the fcc surface is constructed 
from the depicted plaquette in the way that neighboring spins along the $x$ and 
$y$ directions point the same way (alternate) if they are parallel 
(antiparallel) on the plaquette in these two directions. The spins in rows and 
columns where only one spin is presented are continued in the same way as the 
corner spins. For instance in the configuration a) of the (001) surface, both 
the right-hand side and left-hand side neighbors of the ``central'' spin will 
point upwards, while the spin direction will be alternated along the $x$ axis. 
This convention will be maintained henceforth (for a (111) surface one has to 
alter or keep the spins along three axes, instead of two). The smallest set that 
gives a complete idea about the spin structure is presented in Fig.~\ref{fprim} 
\cite{4atoms}; this ``magnetic primitive cell'' does not give a clear picture of 
the crystal symmetries, however. The whole crystal lattice can be reproduced by 
translations of this cell, without performing other operations such as 
reflections or rotations. 

The SHG response types for the (001) monolayer are given in Table~\ref{tab1}, 
for the paramagnetic, ferromagnetic, and all AF phases. We can observe several 
sets of allowed tensor elements. The Conf. r) will produce the same signal as 
the paramagnetic phase. The Conf. ``ferro1'' reveals a completely different, 
distinguishable set of tensor elements. In addition, the conf. ``ferro2'' 
produces another set of tensor elements, different from any other configuration. 
It is equivalent to the conf. ``ferro1'' rotated by 45$^\circ$.
In the confs. a), b), e), and o) we find the 
same tensor elements as for the paramagnetic phase. However, due to the lower 
symmetry, their values are no longer related to each other. Confs. c) and f) 
bring new tensor elements, thus allowing for the distinction of these 
confs. from the previous ones. Confs. i), k), m), p) reveal the same 
tensor elements as c) and f) but some of these elements are related. Thus one 
may possibly distinguish these two sets of configurations. Conf. ``ferro4'' 
presents a completely different, distinguishable set of the nonvanishing tensor 
elements. 
Consequently, in six configurations (i.e. c), f), i), k), m), and p)) some 
susceptibility tensor elements appear only in the AF phase, allowing for the 
detection of this phase by varying the incident light polarization, as will be 
outlined in Sec. IV. In addition, all other antiferromagnetic configurations but 
r) reveal the breakdown of some of the relations between the different tensor 
elements, compared to the paramagnetic phase, and thus can be detected as well. 
Generally, all the phases can be distinguished from each other. There exists as 
well a possibility to distinguish different AF configurations provided the 
corresponding tensor elements can be singled out by the proper choice of the 
experimental geometry. 

\begin{figure}
\epsfig{file=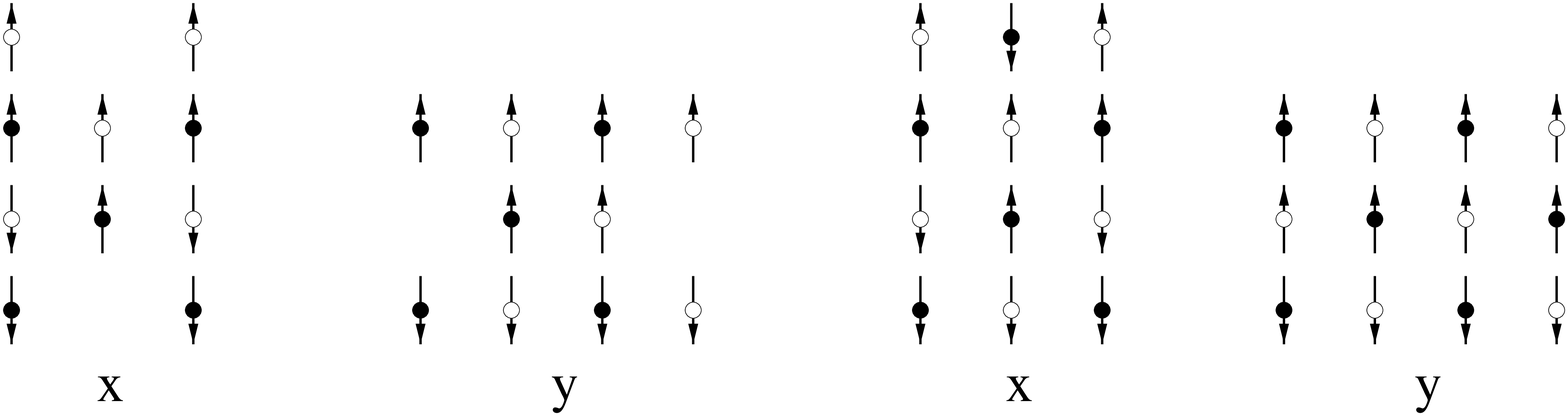, scale = 0.4}
\caption{\label{seclay} Spin structure of an antiferromagnetic (001) bilayer 
constructed from a shift of the monolayer along the positive $x$ ($y$) axis. 
Filled (empty) circles represent the topmost (second) layer. On the right hand 
side the conventional unit cells for the resulting bilayer structure are 
presented. Here, conf. a) of the (001) monolayer serves as an example.}
\end{figure}

\begin{figure}
\epsfig{file=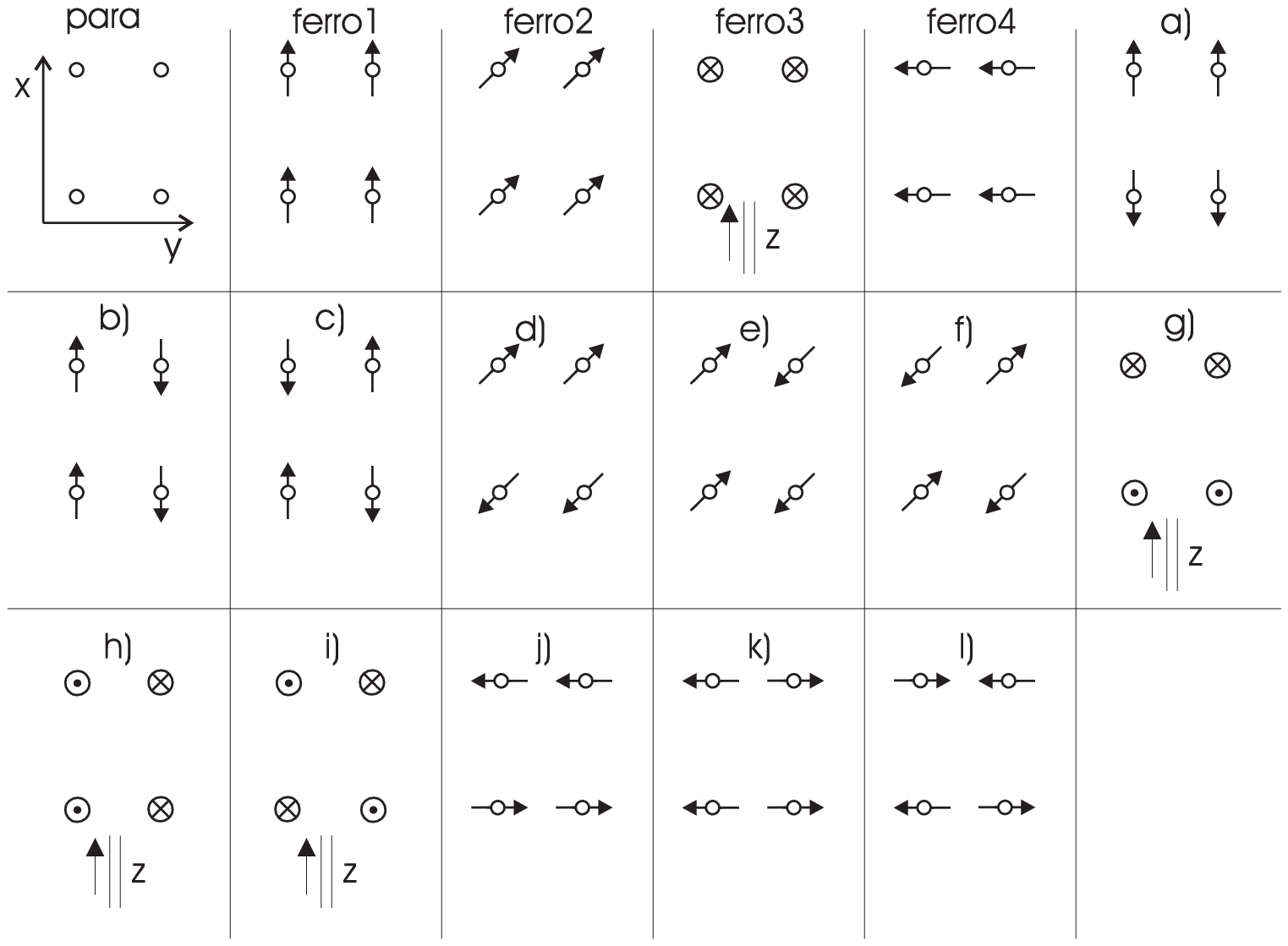, scale=0.9}
\caption{\label{fig2}Spin configurations of an fcc (110) surface. Except for 
confs. ``ferro3", g), h), and i), the arrows always indicate in-plane directions 
of the spins. In confs. ``ferro3", g), h), and i) $\odot$ ($\otimes$) denote 
spins pointing along the positive (negative) z-direction, respectively.}
\end{figure}

\begin{figure}
\epsfig{file=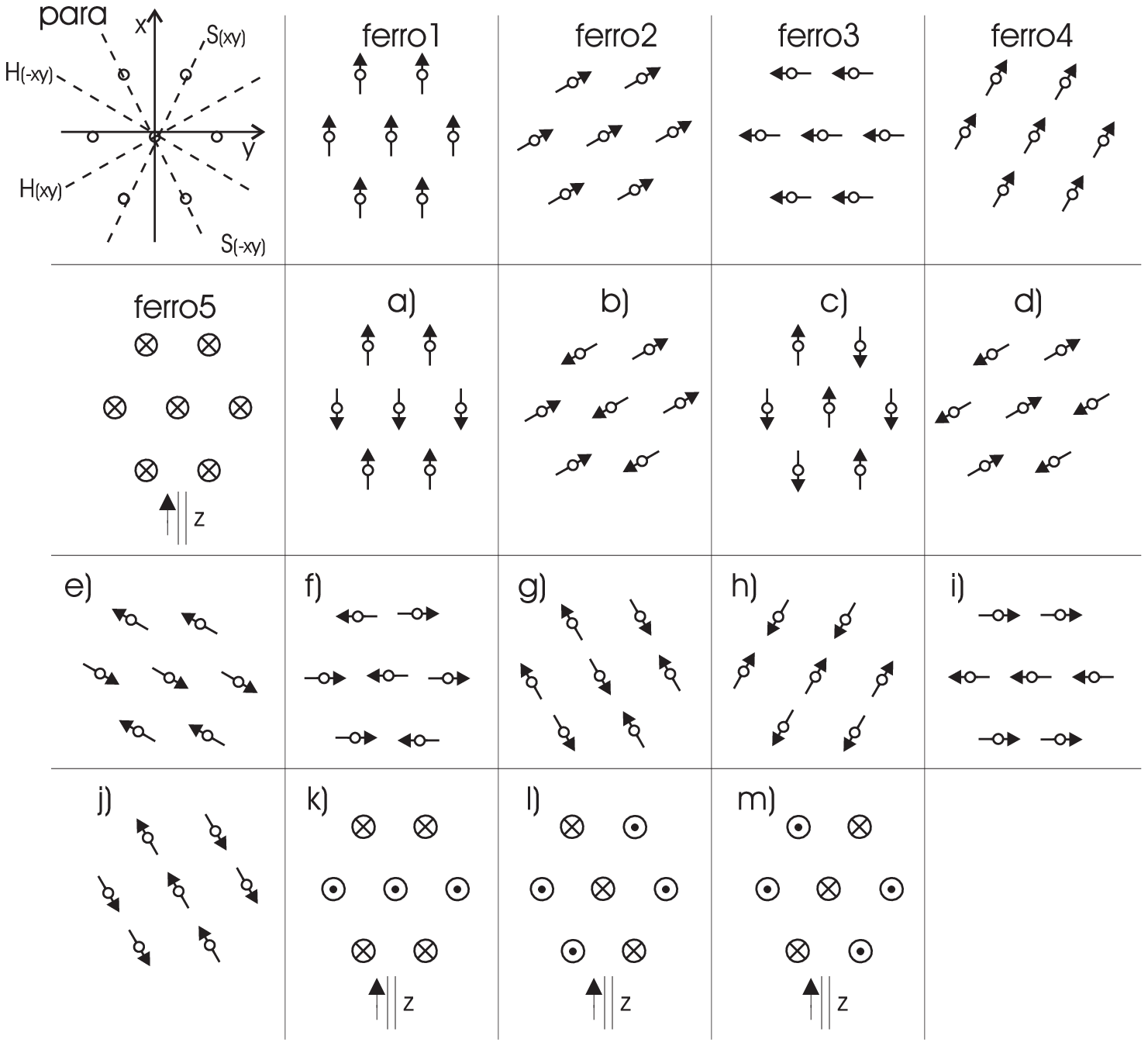, scale=0.95}
\caption{\label{fig3}Spin configurations of an fcc (111) surface. Except for 
confs. ``ferro5", k), l), and m), the arrows always indicate in-plane directions 
of the spins. In confs. ``ferro5", k), l), and m) $\odot$ ($\otimes$) denote 
spins pointing along the positive (negative) z-direction, respectively.}
\end{figure}

For the sake of completeness, we now present a short study of the (001) surface 
where the spin structure of the two topmost atomic layers is taken into account. 
The paramagnetic phase and all the ferromagnetic configurations remain unchanged 
with respect to the results of the previous paragraph (for the (001) monolayer). 
However, most of the AF configurations previously addressed break up into two 
different configurations (sometimes even with a different symmetry). These 
configurations are constructed from the ones of the previous paragraph by 
assuming that the structure of the second atomic layer is identical with that of 
the topmost one but shifted along the positive $x$ axis (indicated by x after 
the name of the original configuration) or positive $y$ axis (indicated by y 
after the name of the ``parent'' configuration) in a proper way to form a fcc 
structure; if only one configuration can be produced in this way we use the name 
of the original one. This construction is depicted in Fig. \ref{seclay}, along 
with the corresponding conventional unit cells for the two topmost layers of the 
AF fcc (001) surface. The resulting SHG response types are presented in Table 
\ref{t002}. In general, seven types of response are possible. Firstly, the 
paramagnetic phase reveals a characteristic set of tensor elements. Thus it can 
be unambiguously distinguished from any other magnetic phase. Secondly, confs. 
``ferro1'', ax), ox), bx), by), ex), and ey) bring some additional tensor 
elements into play. The symmetry of confs. ax) and ox) is slightly different 
from the one of the rest of this group, since the mirror plane is rotated by 
$90^\circ$ around the $z$ axis. A different set of tensor elements is brought up 
by confs. ``ferro2", i), m), and p). The difference between the response yielded 
by conf. i) and the other confs. in this group, due to a slightly different 
symmetry, can be compensated by rotating the sample by $90^\circ$ around the $z$ 
axis. Another, characteristic set of tensor elements is presented by conf. 
``ferro4" alone. The fifth type of SHG response is given by confs. ay), oy), and 
r). Tensor elements, that do not vanish in these configurations, are the same as 
for the paramagnetic phase but some relations between them are broken due to a 
lower symmetry in the AF phase. Confs. cx), fx), and fy) yield all tensor 
elements in an unrelated way. The last, characteristic type of response is 
presented by conf. k) alone. Consequently, the detection possibilities of an 
antiferromagnetic bilayer are slightly worse than those for a monolayer. 
Especially, a difficulty in distinguishing the ferromagnetic phase from the 
antiferromagnetic one may arise for some configurations where then the 
combination of SHG with linear magneto-optics is definitly required. There 
exists a possibility to distinguish AF configurations from each other, similarly 
to the previous situation. In most configurations, the difference (in terms of 
the SHG response) between the bilayer structure described here and the 
previously addressed (001) monolayer can be detected.

We now turn to the (110) surface (Fig.~\ref{fig2}), which, in the paramagnetic 
phase, reveals a lower symmetry than the (001) surface. On the other hand, the 
number of symmetry operations in the AF configurations is comparable to the 
(001) surface. In addition, as shown in Table~\ref{tab2}, the resultin SHG 
response types are not very characteristic, so the detection possibilities for 
this surface are very limited. In particular,  confs. a), b), c), g), h), i), 
j), k), and l) give the same tensor elements as the paramagnetic phase. Confs. 
d), e), f), and ``ferro3'' bring new tensor elements. Other ferromagnetic 
configurations 
(``ferro1'' and ``ferro2'') present different sets of new tensor elements, 
making these configurations distinguishable from the others as well as from each 
other. Conf. ``ferro4'' yields a completely different set of tensor elements, 
however this set is related to the one of conf. ``ferro1'' by $90^\circ$ 
rotation.

The study of the (111) surface (see Fig.~\ref{fig3}) has to be separated in two 
subcases, according to whether we take into account only one atomic monolayer or 
more. In both subcases, we consider the same configurations. The SHG response 
types for the first subcase are listed in Table~\ref{tab3}, and for the second 
subcase in Table 
~\ref{tab4}. For the {\em first} subcase, confs. a), i),  and k) reveal the same
tensor elements as the paramagnetic phase, however due to the  lower symmetry 
their values are not related to each other. Configurations c) and f) present new 
tensor elements. As for the previous surfaces, the ferromagnetic phase reveals  
completely different sets of tensor elements, and the three ferromagnetic 
configurations can be distinguished from each other since they bring different 
tensor elements into play. Unlike for the (110) surface, the axes $x$ and $y$ 
are not topologically equivalent, and thus the fact that tensor elements of 
``ferro1'' are related to those of ``ferro3'' by $90^\circ$ rotation does not 
affect the possibility to distinguish these two configurations. The 
ferromagnetic conf. ``ferro5'' brings up the same tensor elements as AF confs. 
c) and f), but the relations between the elements are different.
The {\em second} subcase (more layers taken into account) gives different sets 
of allowed tensor elements (compared to the first subcase) for each but the
``ferro3'' configuration. Confs. a), i), k), and ``ferro3'' share the same set 
of allowed tensor elements and can be easily distinguished from the paramagnetic 
phase. Confs. c), f), and ``ferro1'' reveal all tensor elements, with their 
values unrelated. Similarly, conf. ``ferro5'' presents another, distinguishable 
set of tensor elements. The possibility to distinguish the magnetic phases is 
rather limited. 

The symmetry analysis of nonvanishing tensor elements for ferromagnetic surfaces 
in the case A have been performed by Pan {\em et. al.} \cite {ref53}. Our 
analysis yields the same results, taking into account the corrections 
made by H\"ubner and Bennemann \cite {ref80}.

\subsection{Distortions of monoatomic lattice}

\begin{figure}
\epsfig{file=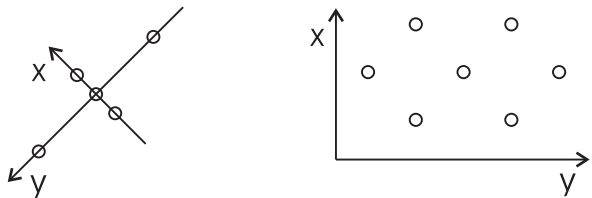}
\caption{\label{fdis111} Structure of the (001) and (111) surfaces of a fcc 
crystal with a rhombohedral distortion in the paramagnetic phase. Note the 
changed orientation of the coordinate system for the (001) surface.}
\end{figure}

\noindent The rhombohedral distortion of the atomic lattice, described here and 
shown in Fig. \ref{fdis111}, makes the $x$ and $y$ axes of the (001) surface 
inequivalent, even in the paramagnetic phase. On the (111) surface, the $y$ axis 
is not equivalent any longer to other axes connecting the nearest neighbors, 
namely $S_{(xy)}$ and $S_{(-xy)}$ (for the definition of 
the ``S'' and ``H'' axes see Fig. \ref{fig3}, the paramagnetic conf.). These 
inequivalences of axes are the reasons for the reduction of the number of 
symmetry operations in the paramagnetic phase. Because of this reduction some 
spin structures that previously formed different domains of a single 
configuration now cannot be transformed into each other and become 
``independent" configurations. This happens for almost every of the previously 
addressed configurations of the (001) and (111) surfaces. Consequently, all the 
depicted spin structures are in fact configurations, and are addressed in this 
subsection. 

The resulting SHG response types for the (001) surface are listed in 
Table \ref {tdis100}. For this surface, only two of the ferromagnetic 
configurations, namely ``ferro1'' and ``ferro2'' can be easily distinguished 
from both the paramagnetic as well as the antiferromagnetic phases. These 
ferromagnetic configurations can be also distinguished from each other. On the 
contrary, all the AF configurations yield only two types of response, and in 
addition one of them is equivalent to the response of the paramagnetic phase. 
Consequently, it will not be possible to determine the surface spin structure, 
and the distinction of the
AF phase from the paramagnetic one can be successfully performed only in confs. 
a)-h) and o). Compared to the case A, there is an important symmetry breaking 
for most configurations. Thus, the distinction between the two cases (A and B) 
is possible (compare Tabs. \ref{tab1} and \ref{tdis100}).

All the (110) surfaces of an fcc crystal with a rhombohedral distortion are 
topographically equivalent to the (110) surface of the case A. The distortion 
only stretches the $x$ or $y$ axis, so the structure remains rectangular. 

The analysis of the (111) surface (depicted in Fig. \ref{fdis111}) in the 
subcase of only one monolayer reveals sets of symmetries very 
similar to the (110) surface, as it follows from the Table \ref{tdis1111}. In 
fact, the (111) surface of a fcc crystal with a rhombohedral distortion can be 
treated as two rectangular lattices superimposed on each other. In turn, due to 
the distortion, it is not convenient any longer to describe the spin structures 
using ``S" and ``H" axes. The possibility to distinguish AF configurations is 
very poor, and two of the AF configurations (a) and k))
yield the same signal as the paramagnetic surface. In confs. b) - j), l), and m) 
the AF phase can be distinguished from the paramagnetic one, but they give the 
same signal as conf ``ferro5". Conf. ``ferro2" can be easily distinguished since 
it reveals a characteristic set of (all) tensor elements. Confs. ``ferro1" and 
``ferro3" yield different sets of tensor elements, but they are related to each 
other by $90^\circ$ rotation. Most of the configurations allow for the 
distinction of the cases A and B (compare Tabs. \ref{tab3} and \ref{tdis1111}).

In the subcase of two monolayers of the (111) surface, the 
symmetry is dramatically reduced (see Tab. \ref{tdis1112}). Even in the 
paramagnetic phase the group of symmetries consists of only one nontrivial 
operation, and this appears to occur also in the AF configurations a), i), k), 
and ``ferro3". In all the other configurations all tensor 
elements are allowed due to the lack of any symmetry. Only confs. paramagnetic 
and ``ferro5" allow for the unambiguous distinction of the cases A 
and B (compare Tabs. \ref{tab4} and \ref{tdis1112}). Consequently, this surface 
is not very useful to an analysis of the magnetic structure, with the exception 
of stating the distortion itself.

As the conclusion of the case of the distorted sublattice of magnetic atoms, the 
surfaces give extremely limited possibilities to 
investigate the magnetic properties. In our further study, we will limit 
ourselves to lattices of undistorted magnetic atoms.

\subsection{Structure with nonequivalent magnetic atoms}

\noindent We assume now that not all the magnetic atoms in the cell are 
equivalent. An example of such a structure is a material composed of two 
magnetic elements, but also a situation when the magnetic lattice sites are 
inequivalent due to different bonds to a nonmagnetic sublattice; distortions of 
the sublattice of nonmagnetic atoms that preserve the center of twodimensional 
inversion produce the same effect. Other distortions of the sublattice of 
nonmagnetic atoms will be 
discussed in subsection D. The magnetic moment at the distinguished positions 
can be changed or not - this does not affect the results obtained by symmetry 
analysis. The configurations considered here are ``ferro1", ``ferro2", 
``ferro4", a), b), c), e), f), i), k), m), o), p), and r) for the (001) surface 
(see Fig. \ref{fig1}), ``ferro1", ferro3", ``ferro5", a), c), f), i), and k) for 
the (111) surface (see Fig. \ref{fig3}), and all configurations depicted in Fig. 
\ref{fig2} for the (110) surface. Other depicted spin structures form domains of 
these configurations and are not referred to in this subsection nor in the 
tables concerning the current subsection.

\begin{figure}
\epsfig{file=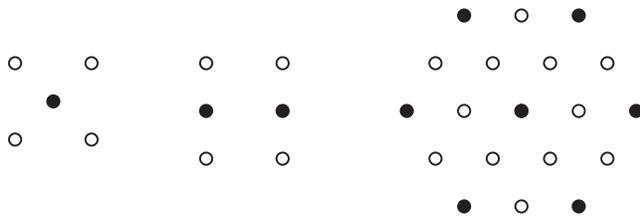}
\caption{\label{fnierow} Surface structure of the non-equivalent magnetic atoms 
case in the paramagnetic phase. Pictures present the (001), (110), and (111) 
surfaces, respectively. Filled and empty circles represent the two kinds of 
magnetic atoms. Note, the fragment representing the (111) surface does not show 
the conventional unit cell but a bigger set of atoms in order to give a clear 
idea about the surface structure.}
\end{figure}

The structure is depicted in Fig.~\ref{fnierow}.
For the sake of brevity, we show the structure of the distinguished atoms only 
for the paramagnetic phase. All the configurations are the same as in case 
A, for all surface orientations. The already mentioned ``convention'' of 
alternating (or not) spin directions along certain axes 
is applied regardless of the atom type. This allows us to obtain the whole 
crystal surface from the small displayed fragment.

Our analysis starts with the (001) surface of an fcc crystal. The
SHG response types for each configuration are listed in Table~\ref{tnierow100}.
In general, we can observe seven types of response. The first of them is 
represented by the paramagnetic phase alone. The second type of response, 
exhibited by the ferromagnetic ``ferro1'' and the 
AF a), b), e), o) confs., differs from any other type by some tensor elements.
The confs. a) and o) reveal different tensor elements than the other 
configurations from the mentioned group. However, the signal from confs. a) and 
o) is the same as for the confs. b), e), and ``ferro1'' if one exchanges the 
axes $x$ and $y$. Thus, 
if the directions of the spins cannot be determined by another method, 
confs. a) and o) cannot be distinguished from b), e), and ``ferro1''. The next 
type consists of conf. f) and reveals all tensor elements, while 
no relations between them are enforced by the symmetry analysis. 
A completely different type of response is presented by conf. c) alone. 
Another type, where confs. i), m) and p) belong to brings the same tensor 
elements as conf. c), but there exist more relations between the elements due to 
a higher symmetry in these configurations. The next type is given by confs. 
``ferro2'' and k). As in conf. f) all the tensor elements are present 
but this time there are some relations between them. In addition, confs. r) and 
``ferro4'' yield a completely new set of tensor elements due to the preserved 
fourfold rotational symmetry.

Thus, assuming one atom as distinguished may reduce the symmetry. New 
tensor elements appear in confs. a), b), e), f), k), o), and r) compared to  
case A (compare Tabs. \ref{tab1} and \ref{tnierow100}). In these configurations 
it is therefore possible to distinguish the cases of equivalent and 
nonequivalent magnetic atoms, provided the 
tensor elements that make the cases different can by singled out by the 
experimental geometry. There exists also a possibility to distinguish different 
AF configurations in case C. The antiferromagnetic {\em phase} can be 
undoubtelly detected in the surface configurations c), f), i), m), and  p).

For the (110) surface, there are more possibilities to distinguish the 
configurations with nonequivalent magnetic atoms than in the case A. However, 
the configurations still produce ambiguous signals (see Tab. \ref{tnierow110}). 
Confs. b), c), h), i), k), and l) are equivalent to the paramagnetic phase. 
Conf. a) is equivalent to the ferromagnetic ``ferro1'' configuration, and 
conf. d) to ``ferro2''. In addition, the confs. e), f), and g) are equivalent to 
the conf. ``ferro3'' and conf. j) gives the same signal as conf. ``ferro4''. 
Even the presence of nonequivalent atomic sites in the lattice cannot be 
detected by SHG on this surface, since the symmetry of the (110) surface is 
usually not lowered further by the existence of equivalent magnetic sites 
(compare Tables \ref{tab2} and \ref{tnierow110}. The only exception are the 
confs. a), d), g), and j) which give different tensor elements in the two cases. 
As in the case of equivalent atoms, the (110) surface is not very useful for the 
analysis.

The study of the (111) surface must again be divided in the two subcases of one 
or more monolayers, respectively. Fig.~\ref{fnierow} depicts the situation in 
the paramagnetic phase. The SHG response types are listed in Tables 
~\ref{tnierow1111} and~\ref{tnierow1112} for the first and the second subcase 
respectively. 

In the first subcase (one monolayer) the symmetry establishes six different 
types of nonlinear response. The ``paramagnetic'' type (for the paramagnetic 
configuration only) is characteristic - all the other configurations have 
additional tensor elements. The next type of response (the ferromagnetic conf. 
``ferro1'' and the antiferromagnetic conf. a)) brings some new tensor elements. 
Other tensor elements appear in the conf. k). Configurations ``ferro3'' and i) 
show another set of nonvanishing tensor elements. The confs. c) and f) reveal 
all tensor elements in an unrelated way. In addition, conf. ``ferro5'' presents 
a characteristic set of tensor elements.

In the second subcase, only four different SHG responses are possible. Firstly, 
the paramagnetic phase is characteristic - all the other configurations bring 
additional tensor elements into play. The next type of response is presented by 
confs. ``ferro3'' and i) - they yield some additional tensor elements. Confs. 
``ferro1'', a), c), f), and k) reveal all tensor elements and no relations 
between them appear from our symmetry analysis. Again, the conf. ``ferro5'' 
presents a unique set of nonvanishing tensor elements.

Consequently, for the (111) surface, the symmetry breaking due to the presence 
of a second kind of magnetic atoms has even more important consequences than for 
the (001) surface. In the situation of only one monolayer, the distinction 
between the cases may be possible for all the AF configurations (compare Tables 
\ref{tab3} and \ref{tnierow1111}). Considering additional layers leads to 
further symmetry breaking and renders the distinction between the configurations 
impossible. The distinction between the cases A and C is possible in confs. a) 
and k) (compare Tables \ref{tab4} and \ref{tnierow1112}). Besides, in most 
configurations it is possible to decide if these additional layers play any 
role (compare Tables \ref{tnierow1111} and \ref{tnierow1112}).

\subsection{Distorted oxygen sublattice}

\noindent Due to the strong charge-transfer between nickel and oxygen in NiO the 
sublattices may be distorted. This effect can lower the symmetry of the surface. 
A point-charge model calculation by Iguchi and Nakatsugawa \cite{ref42} 
presented a shift of the oxygen sublattice (``rumpling'') in the direction 
perpendicular to the surface. Their method did not show any in-plane 
displacement and thus no change of the surface symmetry. However, if the 
``rumpling'' also has an in-plane component, i.e. if the oxygen atoms are 
displaced also in the $x$ and $y$ directions, it will also have a considerable 
effect on the symmetry of the crystal surface. For this paper, we have chosen a 
distortion that can lower the symmetry of the surface and besides can be 
represented within one conventional unit cell. The configurations considered 
here are ``ferro1", ``ferro2", ``ferro4", a), b), c), e), f), i), k), m), o), 
p), and r) for the (001) surface (see Fig. \ref{fig1}), ``ferro1", ferro3", 
``ferro5", a), c), f), i), and k) for the (111) surface (see Fig. \ref{fig3}), 
and all configurations depicted in Fig. \ref{fig2} for the (110) surface. Other 
depicted spin structures form domains of these configurations and are not 
referred to in this subsection nor in the tables concerning the current 
subsection.

As will be shown later, the best conditions for the detection of this kind 
of distortion are presented by the (110) surface. The (111) surface could show 
equally good possibilities if only a monolayer of magnetic atoms is present.

In the presence of an oxygen sublattice distortion, the chemical unit cell is 
also doubled. This effectively means that magnetic unit-cell-doubling 
(describing the fact that the magnetic unit cell is twice as big as the chemical 
one) is lifted. In general, taking into account distorted oxygen atoms in the
paramagnetic phase does not lower the symmetry of the problem. The exception is 
the (111) surface, where the six-fold axis is replaced by the three-fold one. 

\begin{figure}
\epsfig{file=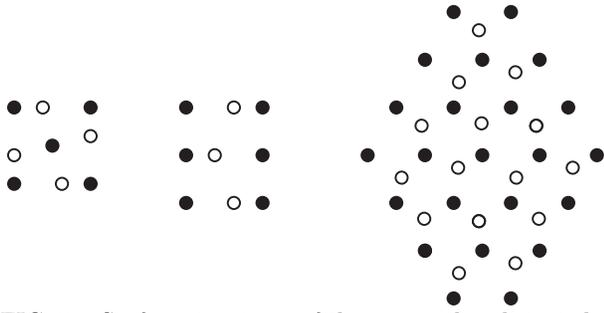}
\caption{\label{ftlen} Surface structures of the case with a distorted oxygen 
sublattice (white circles). Pictures present the paramagnetic phase of (001), 
(110), and (111) surfaces, respectively. Note, the fragment representing the 
(111) surface does not show the conventional unit cell but a bigger set of atoms 
in order to give a clear idea about the surface structure.}
\end{figure}

In the case of the distorted oxygen sublattice, the symmetry group for each 
configuration is a subgroup of the corresponding ``non-distorted'' 
configuration, i.e. of the corresponding spin configuration in the case where 
the oxygen atoms are not considered. As in case C we display only the 
paramagnetic phase in Fig. \ref {ftlen} to depict the atom positions. All the 
spin configurations are the same as for the corresponding surfaces in case A, 
and the spins are assumed to be equivalent.

As Table~\ref {ttlen100} shows, six different responses can be expected 
from the (001) surface. The paramagnetic surface will give a characteristic 
response. The second group is formed by the confs.: a), b), e), o), and 
``ferro1''. Although confs. a) and o) have elements different from the remaining 
configurations in this group, this fact corresponds simply to rotating the 
sample by 90$^\circ$ with respect to the $z$ axis. Confs. c) and f) reveal 
all tensor elements without relations between them. Confs. ``ferro2'', i), k), 
and m) reveal all tensor elements with some relations. The only difference 
between conf. m) and others from this group is like for the previous group a  
90$^\circ$ rotation with respect to the $z$ axis. Another group consists of 
conf. p) alone. It reveals the same tensor elements as the paramagnetic phase, 
but certain relations between tensor elements are broken due to a lower symmetry 
of the conf. p). The confs. r) and ``ferro3'' form the last group. All the 
configurations but k) and ``ferro3'' can be distinguished from those of case A 
(compare Tabs. \ref{tab1} and \ref{ttlen100}). However only confs. c) and g) can 
be distinguished from case C (compare Tabs. \ref{tnierow100} and 
\ref{ttlen100}). Thus, only in these configurations it will be possible to 
detect oxygen sublattice distortions by SHG.

The SHG response types for the (110) surface are presented in Table 
\ref {ttlen110}. One can observe that only configurations c), f) and i) give 
rise to new (compared to case A, Table \ref{tab2}) tensor elements. Compared to 
case C (Table \ref{tnierow110}), 
confs. c), f), and i) bring new tensor elements, and, surprisingly, confs. a) 
and g) have less tensor elements, due to higher symmetries in the 
case D. Consequently, the confs. a), c), f), g), and i) allow for an unambiguous 
determination of the oxygen sublattice distortion from the (110) surface. The 
possibility to distinguish different configurations is rather limited.

Oxygen sublattice distortion similar to the one presented in Fig. \ref{ftlen} 
for a (111) surface was found by Renaud {\em et al.} \cite{ref68} and calculated 
by Gillan \cite{refGillan} in M$_2$O$_3$ materials (M = Al, Fe). Since the 
nonmagnetic sublattice symmetry group has an influence on SHG this distortion 
can be detected also on surfaces of fcc crystals. In the previous cases A and C 
we divided the study of (111) surfaces in two subcases, considering
either one or more atomic layers. Taking into account a distorted oxygen 
sublattice leads us immediately to the subcase of ``more atomic layers''. It is 
caused by the fact that the oxygen and magnetic atoms belong to mutually 
exclusive planes. The resulting SHG response types are listed in Table \ref 
{ttlen111}. For the AF and ferromagnetic phases, all tensor elements are allowed 
for every configuration.  Thus SHG cannot detect the magnetic phase of the 
surface nor distinguish different configurations. Only confs. paramagnetic, 
``ferro3'', ``ferro5", and d) allow to 
decide unambiguously whether the oxygen sublattice is distorted or not 
(compare Tabs. \ref{tab4}, \ref{tnierow1112}, and \ref{ttlen111}). 

For both the (001) and (111) surfaces, the symmetry groups of case D appear to 
be the subgroups of the corresponding configurations of case C. This means that 
the oxygen sublattice distortion makes some (one half of all) magnetic atoms 
distinguished as in case C, even though we did not apply this distinction 
explicitly in case D. On the other hand, the symmetry groups of the case D 
differ essentially from those of case B. This is caused by the difference in 
distortions assumed in these cases: the rhombohedral one in case B and 
rotation-like in case D.

\subsection{Domain imaging}

\noindent For simplicity, we will consider here only surfaces described hitherto 
by the case A of our analysis. In this case, for AF surfaces, no $180^\circ$ 
domains can be expected due to the presence of magnetic unit-cell doubling. The 
allowed domains can be detected by surface-sensitive SHG under the following two 
conditions.

First, domains can be imaged by our method only if they manifest 
themselves at the surface, i.e. if the surface spin ordering changes while 
passing from one domain to another \cite {domeny}. It is necessary to note, 
however, that the spin orderings for different domains must belong to the same 
{\em configuration} in 
the sense of our classification. We do not consider it as a domain structure if 
one portion of the surface is in one configuration and another portion is in a 
different configuration. Under such conditions, we can encounter two different 
types of domains: $90^\circ$ domains (for the (111) surface they are rather 
$60^\circ$ domains), resulting from the rotations around $z$ 
axis, and the second type (called by us mirror-domains, characteristic for 
antiferromagnets), where spins point along the same axis in all domains, 
but the ordering is still different (they are no $180^\circ$ domains!). The 
tables contain the complete information about the parity of tensor elements in 
mirror-domain operations, and also for $90^\circ$ type domains, but not for  
$60^\circ$ domains. The $90^\circ$ type domains will be addressed later on.
In the mirror-domain structure, the magnetic point group 
describing the configuration must lack an operation that, while belonging to the 
(nonmagnetic) point group of the system {\em and} leaving the spin axes 
invariant, only flips some of the spins. 
Note, the flipped subset of the spins must be antiferromagnetically ordered in 
itself. Configurations, the symmetry groups of which {\em lack} one of these 
operations can reveal surface domains, related to each other by this operation. 

\begin{figure}
\epsfig{file=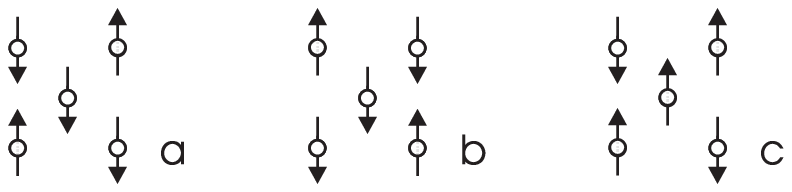}
\caption{\label{fimage} Two surface mirror domains for an AF configuration - 
panels b) and c) depict the same AF domain, related to the panel a) by different 
mirror operations.}
\end{figure}

For an illustration we choose the configuration c) of the (001) surface (see 
Fig. \ref{fig1}). The spins point along the $x$ axis. Thus operations leaving 
the axis invariant are $\overline 2_x$, $\overline 2_y$ and $2_z$. Of them, 
$\overline 2_x$ and $\overline 2_y$ are absent in 
the magnetic point group of the considered configuration (see Tab. \ref{tab1}, 
conf. c), and Tab. \ref{keytab}). The flipped subset of spins consists of the 
four outer spins for the $\overline 2_x$ operation, and of the central spin for 
$\overline 2_y$ (see Fig. \ref{fimage} b) and c), respectively). In fact, there 
are two 
domains possible in this configuration: one with the spins kept invariant under 
translations by the vector ($-{a\over 2},{a\over 2},0$) (this domain is shown) 
and the other with the spins kept invariant under translations by the vector 
(${a\over 2},{a\over 2},0$). Here, $a$ denotes the lattice constant. These 
domains are depicted in Fig. \ref{fimage}. 

The second condition for domain imaging is an interference. It can be 
created internally by different elements of the tensor $\chi^{(2\omega)}$ or by 
external reference \cite {external1,external2}. The interfering elements should 
be of a similar magnitude for the largest possible image contrast. Group theory, 
however cannot account for the amplitudes. With external as well as internal 
reference, a tensor element that changes its sign under the reversal 
of the antiferromagnetic order parameter {\bf L} is necessary. Actually, every 
{\bf L} 
dependence of $\chi^{(2\omega)}$ can be represented by splitting the tensor 
elements into odd and even ones in {\bf L}; even if a tensor element is not 
purely 
odd or even we can always decompose it according to:
\begin {equation}
\label{element}
\chi^{(2\omega)}_{ijk}=\chi^{(2\omega ),odd}_{ijk}+\chi^{(2\omega),even}_{ijk}
\end{equation}
i.e. a tensor element consists of parts which are odd and even in {\bf L}, 
respectively. In a system with 
many terms of that kind the possibility of detecting domains may be limited, 
since they can influence the signal with opposite sign, thus diminishing the 
interference. In highly symmetric structures, such as an fcc crystal, the 
situation is more comfortable: every tensor element is either odd or even in 
{\bf L} (see Appendix). By the appropriate set of experiments an element can be 
singled out and give a clear image of AF domains.

As an example we consider tensor elements that are present in all the
phases, e.g. $\chi^{(2\omega)}_{zzz}$: they are even in the magnetic 
order parameters {\bf L} and {\bf M}, for the AF and ferromagnetic phases, 
respectively. The tensor element $\chi^{(2\omega)}_{zxy}$, present for example 
in the previously discussed conf c) of the (001) surface (see Fig. 
\ref{fimage}), is odd, since it changes its sign under the operation 
$\overline{2}_x$ transforming one domain into another. For other configurations 
other 
tensor elements and operations can be found. In the discussed configuration both 
these elements are present, we have intensity contributions proportional to 
$(\chi^{(2\omega)}_{zzz})^2$, $(\chi^{(2\omega)}_{zxy})^2$ and 
$\chi^{(2\omega)}_{zzz}\cdot 
\chi^{(2\omega)}_{zxy}$, due to the square in Eq. (\ref{eqGlown}). As a 
result, one obtains an interference:
\begin{equation}
I_p\sim ... +(\chi^{(2\omega)}_{zzz})^2+(\chi^{(2\omega)}_{zxy})^2\pm 
2\chi^{(2\omega)}_{zzz}\cdot \chi^{(2\omega)}_{zxy}+...
\end{equation}
where ``+'' stands for one domain, ``-'' for a different one.

Now, we turn to the $90^\circ$ domain structure. Again, we take the conf. c) of 
the (001) surface as an example. The operation connecting the domains is $4_z$. 
Under this operation, the tensor element $\chi^{(2\omega)}_{zxy}$ changes its 
sign, thus again we have an interference which renders the domain imaging 
possible. This tensor element is even in the domain operation $\overline 2_{xy}$ 
(which is equivalent to the superposition of $\overline 2_x$ and $4_z$), which 
means that domains related to each other by this operation cannot be imaged 
using this particular tensor element. Similarly, if a tensor element is odd in 
one domain operation and even in another, it must be odd in their superposition. 
Concerning the $60^\circ$ domains for (111) surfaces, the parity of the tensor 
elements must be treated more carefully, as indicated already in \cite{par111}. 
We can still define three ``twofold'' operations, and each of them has its own 
set of odd and even tensor elements. The sets corresponding to different of 
those operations are not mutually exclusive, i.e. a tensor element is usually 
shared among different parities. In this way, this tensor can be positive in one 
domain, negative in the second, and zero in the third one. Thus, the existence 
of a well defined parity of tensor elements is necessary for domain imaging, but 
not sufficient for $60^\circ$ and $120^\circ$ domain structure.

This unleashes an interesting question of the antiferromagnetic order parameter. 
There are as many order parameters as different domain structures for a given 
configuration. For $60^\circ$ and $120^\circ$ domain structures, the AF order 
parameter must be a vector, while for mirror domains it is a number. For 
$90^\circ$ domains it is can be also a number, since there are only {\em two} 
$90^\circ$ domains. The vectorial order parameter transforms itself under the 
domain operation like a usual vector.

It is necessary to mention at this point that taking into account the spin 
structure in the {\em second layer} would not change the validity of the 
analysis presented in this subsection. The only modifications would result from 
addressing bulk domains rather then surface domains, and the symmetry of the AF 
configurations would be changed. Yet it would still be possible to find domain 
operations as well as odd and even tensor elements leading to  interference and 
AF domain contrast. However the possibility to identify each of the domains may 
be limited in some cases due to the increased number of domains.

\section{Possible experimental setups}
\begin{figure}
\epsfig{file=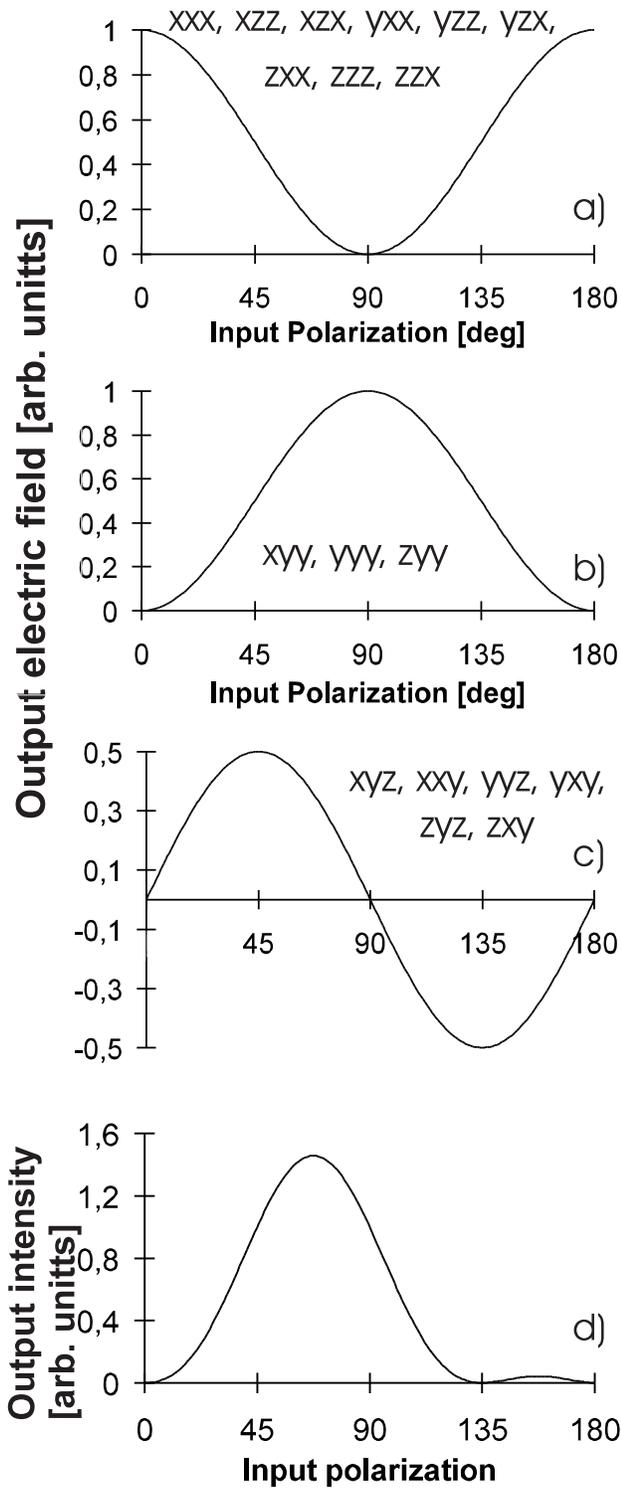}
\caption{\label{fEl} Electric field response of single tensor elements as a 
function of the input polarization. Tensor element $\chi^{(2\omega)}_{ijk}$ is 
denoted as ijk. Graph d) shows an example of the SHG light intensity.}
\end{figure}

\noindent
In this section, we propose and discuss possible experimental setups for the 
detection of AF configuration and the imaging of AF domains from low-index 
surfaces of NiO that exhibit magnetic unit-cell doubling in contrast to bulk 
$Cr_2O_3$ \cite {ref18,ref4}.
We propose an experimental setup for the {\em detection} of antiferromagnetism 
in the following way: both the incident 
and reflected beams may lie in the $xz$ plane (optical plane), and form the 
angle $\vartheta$ with the $z$-axis (normal to the sample surface). In the plane 
perpendicular to the outgoing beam axis, the electric 
field of the second-harmonic generated light has two 
components, $E^{(2\omega)}_p$ and $E^{(2\omega)}_s$, given by the formulae
\begin{equation}
\begin{split}
\label{electric}
|E^{(2\omega)}_p| &=|\cos \vartheta E^{(2\omega)}_x-\sin \vartheta 
E^{(2\omega)}_z| \\
|E^{(2\omega)}_s| &=|E^{(2\omega)}_y|
\end{split}
\end{equation}
$E^{(2\omega)}_x$, $E^{(2\omega)}_y$, and $E^{(2\omega)}_z$ are the components 
of the electric field resulting from SHG in the coordinate system of the sample. 
The dependence of these components on the input electric field is indicated by 
the tensor $\chi^{(2\omega)}$. The aim of the experiment is the determination of 
vanishing and nonvanishing tensor elements. The easiest way to do this is to 
analyze the output signal intensity as a function of the input polarization in 
both output polarizations $s$ and $p$, for a fixed angle of incidence and 
reflection. The dependence of the output second-harmonic 
electric field on the input polarization is schematically displayed
in Fig. \ref{fEl}a)-c) for all tensor elements. The intensity of SHG light is 
the square of the linear combination of these partial responses. An example of 
the intensity dependence on the input polarization is presented in Fig. 
\ref{fEl}d). The intensity need not be symmetric with respect to $\varphi = 
90^{\circ}$, this results from the influence of the electric field depicted in 
Fig. \ref{fEl}c). The coefficients of the mentioned combination are the products 
of the $\chi^{(2\omega)}$ 
tensor elements and the corresponding Fresnel coefficients, according to Eq.~ 
(\ref{eqGlown}). Thus performing a best fit of these coefficients to the 
experimental results will give (after taking into account the Fresnel and 
geometrical
coefficients, known for the given experimental geometry and material 
\cite{Sipe}) a set of 
non-vanishing elements of the $\chi^{(2\omega)}$ tensor. Thus for instance, the 
magnetic phase can be determined.

Concerning another experimental geometry, with input polarization fixed and 
intensity measured as a function of the output polarization, it is possible to 
determine whether the nonlinear Kerr effect takes place. For instance, with the 
input polarization $\varphi=90^\circ$, the output electric field is given as 
follows \cite{Sipe}:
\begin{multline}
\label{varout}
E^{(2\omega )}=\sin \Phi (A_2(\Theta)\chi^{(2\omega)}_{yyy}B_2(\vartheta))+\\
\cos \Phi (A_1(\Theta)\chi^{(2\omega)}_{xyy}B_2(\vartheta) +
A_3(\Theta)\chi^{(2\omega)}_{zyy}B_2(\vartheta))
\end{multline}
As the result, maximum of the intensity is for $\Phi \neq 90^\circ$, if at least 
one of the tensor elements $\chi^{(2\omega)}_{xyy}$ or $\chi^{(2\omega)}_{zyy}$ 
does not vanish. Actually, tensor element $\chi^{(2\omega)}_{zyy}$ is even in 
all the investigated order parameters, but the tensor element 
$\chi^{(2\omega)}_{xyy}$ can be odd. For such configurations the Kerr effect 
(change of polarization caused by inversion of the magnetic order parameter) 
takes place. Thus, it is possible to determine which tensor elements are 
associated with the spin-orbit coupling.

The geometry with p polarization of the reflected SHG light seems to be less 
useful, since there the tensor element 
$\chi^{(2\omega)}_{zzz}$ is always present, regardless of the configuration. 
Besides, this polarization mixes the $\chi^{(2\omega)}_{x..}$ and 
$\chi^{(2\omega)}_{z..}$ tensor elements. This mixing, however, can be tuned by 
varying 
the angle of incidence $\vartheta$ and taking into account the influence of the 
Fresnel coefficients. For smaller $\vartheta$ only the $\chi^{(2\omega)}_{x..}$ 
elements are important, while for larger $\vartheta$ the 
$\chi^{(2\omega)}_{z..}$ 
dominate. If the experiment does not show any difference for 
these two situations, the tensor elements must be related. This is the 
possibility to distinguish the configurations with some relations between the 
tensor elements from those without such relations. On the other hand, the p 
polarization is useful for AF domain {\em imaging}. Thus one of the
experimental possibilities is to carry out the measurements first in s polarized 
outgoing SHG light to make sure that the material is in the AF phase and 
determine its spin configuration. Then a second measurement in p polarization 
can be performed for the domain imaging.

\section{Conclusions}
\noindent
Already a short look at the presented tables shows that our method works best if 
the paramagnetic phase is of high symmetry, since then a wide variety of 
different symmetries exists which may be broken by different spin 
configurations. In other words, there is enough room for different new tensor 
elements to appear along with different 
spin ordering under these circumstances. In general, this is the main reason why 
only nonlinear optics is suited for the detection of antiferromagnetism 
and the imaging of AF domains. The linear susceptibility tensor has too low a 
number of elements for these purposes in order to produce unambiguous results. 
Similarly, among the considered surfaces, the (110) surface is the 
least useful for the analysis as it yields ambiguous signal interpretations due 
to its low symmetry in the paramagnetic phase, and, on the other hand, very 
similar symmetries in all the AF configurations. 

The (001) and (111) surfaces present alike possibilities of distinction between 
the cases. If more than one monolayer is involved, however, the (111) surface 
will give the same response in the cases A (all atoms equivalent) and C (two 
kinds of magnetic atoms). Both the (001) and (111) surfaces also allow for the 
determination of the spin structure, provided the case is known. The (111) 
surface in the case D (oxygen sublattice distortion) is an exception - all the 
AF configurations produce the same response. It is possible, however, to 
determine the phase of the material.

The case D appears to be a subgroup of the case C, i.e. all the magnetic point 
groups describing the configurations of the case D are subgroups of the 
corresponding ones in the case C. The only exception is the (110) surface. This 
inclusion means that the oxygen sublattice distortion makes some (one half of 
all) magnetic atoms distinguished as in case C, even though we did not apply 
this distinction explicitly in case D.

From the fact, that the influence of oxygen sublattice distortion (case D) is 
not detectable in the paramagnetic and ferromagnetic phases it follows that only 
antiferromagnetic ordering can give an extensive information about the structure 
of the surface. It is the magnetic atoms and their magnetism which reveal the 
presence and position of oxygen. 

Our short analysis of an AF bilayer structure (surface (001)) indicates very 
similar features to the (001) monolayer. There exists a possibility to 
distinguish AF configurations from each other, and a certain possibility to 
detect the magnetic phases. Furthermore, introducing the second atomic layer 
does not affect the possibility to image AF domains.

Concerning the {\em magnetic} phases, configurations and cases considered in 
this paper, some {\em a priori} information about the {\em structure} is needed 
in order to draw unique conclusions from the experimental results. For the 
detection of the phase and the spin configuration this additional information is 
the case (A, B, C, D). Vice versa, the case (for instance a possible distortion 
of the oxygen sublattice) can be determined if 
one knows the configuration (and if it had been previously deduced that the 
investigated material is antiferromagnetic). Actually, in most measurements of 
AF spin structures some {\em a priori} knowledge is required. For example, in 
experiments by Fiebig {\em et. al.} \cite{ref4} such a prerequisite is the 
assumption of the AF spin-flop phase of the material. In both experimental 
approaches mentioned here the (001) surface seems to provide the best 
possibilities of drawing valuable conclusions, while the (110) surface is the 
least suitable in that respect.

Finally, our paper demonstrates that the AF domain imaging is possible even in 
the presence of magnetic unit-cell doubling. Thus optical SHG, unlike linear 
optics, is able to image AF {\em surface} domains. For most AF configurations, 
there are more than one surface domain structures. The rule stating that the 
number of domains is equal to the number of symmetry operations in the 
paramagnetic phase divided by the number of symmetry operations in the magnetic 
phase is applicable also for antiferromagnets (thus, with unit-cell doubling the 
number of domains is reduced by a half). However, not all the domains can be 
imaged at the same time.

\section*{ Appendix \\ On the group-theoretical analysis of magnetic systems}
\noindent 
In this Appendix, we would like to address some particularities of our 
group-theoretical analysis. The first general remark is that although symmetry 
analysis can provide us with a set of nonvanishing tensor elements for a given 
configuration, but cannot give any information about their magnitude. This 
equally applies to the distortion effects, as treated e.g. in Ref. \cite{ref56}.

Another interesting issue is the behavior of the tensor elements with respect to 
the AF order parameter {\bf L} (for ferromagnetic phases {\bf L} should be 
replaced by the magnetization {\bf M}), i.e. the parity of tensor elements. In 
general, a tensor element consists of 
even and odd parts with respect to {\bf L}, as shown in Eq. (\ref{element}). In 
systems with high symmetry, it is possible to describe an operation which 
reverses {\bf L} (or {\bf M}) by a spatial operation $\Hat{l}$. The operation 
$\Hat{l}$ belongs to the point group of the system, but not to 
its magnetic point group. The application of this operation to a tensor element 
will change its sign (keep it invariant) if this element is odd (even) in {\bf 
L}. Consequently, each tensor element can be either odd or even in {\bf L}, a 
mixed behavior is forbidden. Actually, the parity of a given tensor element is a 
function of the chosen operation $\Hat{l}$. In most antiferromagnetic 
configurations more than one operation leading to different domain structures 
are possible (this means, more than one order parameter can be defined). For 
example, for (001) surface one has $4_z$ rotations leading to different domains 
{\em in addition} to the eventual mirror-structure. For the (111) surface, there 
are three domains resulting from the rotations with respect to the $z$ axis 
alone. For some configurations, they exist in addition to the mirror-domains.

This whole analysis of the parity of the tensor elements cannot be performed for 
the systems with a lower symmetry, where it is impossible to find an operation 
$\Hat{l}$ describing the inversion of {\bf L} or {\bf M}, a mixed behavior is 
then allowed. Note, the presence of dissipation (redistribution of response 
frequencies) does not influence the above consideration. In general, dissipation 
in frequency space is responsible for the mixing of the real and imaginary parts 
in the tensor elements, while point-group symmetry governs the (non)existence of 
tensor elements purely odd or even in the magnetic order parameters {\bf L} or 
{\bf M}.

We acknowledge financial support by TMR network through NOMOKE contract no. 
FMRX-CT96-0015. We also acknowledge numerous fruitful discussions with Dr. R. 
Vollmer.

\end{multicols}
\newpage

\newpage

\begin{table}
\caption[keytab]{\label{keytab} Details of SHG response types. We denote 
$\chi^{(2\omega)}_{ijk}$ by ijk. Odd elements are in bold if a domain operation 
exists.}
\begin{tabular}{cclcl}
key&point group&symmetry operations&domain operation&non-vanishing 
tensor elements\\ \hline
&&&&\\
a&4mm&$1,2_z,\pm4_z,\overline{2}_x,\overline{2}_y,\overline{2}_{xy},
\overline{2}_{-xy}$&-&xxz = xzx = yyz = yzy, zxx = zyy, zzz\\
b&m&$1,\overline{2}_x$&$2_z,\overline{2}_y$& xzx = xxz, {\bf xxy} = {\bf 
xyx}, {\bf yxx}, {\bf yyy}, {\bf yzz}, \\
& & && yyz = yzy, zxx, zyy, zzz, {\bf zyz} = {\bf zzy} \\
&&&$4_z,\overline{2}_{xy}$& no information about the parity \\
c&m&$1,\overline{2}_{xy}$&$2_z,\overline{2}_{-xy}$& {\bf xxx} = -{\bf yyy}, 
{\bf xyy} = -{\bf yxx}, {\bf xzz} = -{\bf yzz}, \\
& & & & xyz = yxz = xzy = yzx, xxz = xzx = yyz = yzy, \\
& & & & {\bf xxy} = -{\bf yyx} = {\bf xyx} = -{\bf yxy}, zxx = zyy, zzz, \\
& & & & {\bf zxz} = {\bf zzx} = -{\bf zyz} = -{\bf zzy}, zxy = zyx \\ 
&&&$4_z,\overline{2}_y$& {\pmb {\em xxx}} = {\em -yyy}, xyy = yxx, xzz = -yzz,\\
&&&& {\bf xyz} = {\bf xzy} = {\bf yxz} = {\bf yzx}, xxz = xzx = yyz = yzy, \\
&&&& {\em xxy} = -{\pmb {\em yyx}} = {\em xyx} = -{\pmb {\em yxy}}, zxx = zyy, 
zzz, \\
&&&& {\pmb {\em zxz}} = {\pmb {\em zzx}} = {\em zyz} = {\em zzy}, {\bf zxy} = 
{\bf zyx} \\
d&4&$1,2_z,\pm4_z$&$\overline{2}_x,\overline{2}_y,\overline{2}_{xy},
\overline{2}_{-xy}$& {\bf xyz} = {\bf xzy} = -{\bf yxz} = -{\bf yzx}, \\
& & & & xzx = xxz = yzy = yyz, zxx = zyy, zzz \\
e&mm2&$1,2_z,\overline{2}_x,\overline{2}_y$&$\pm 
4_z,\overline{2}_{xy},\overline{2}_{-xy}$& xxz = xzx, yyz = yzy, zxx, zyy, zzz\\
f&2&$1,2_z$&$\overline{2}_x,\overline{2}_y$& {\bf xyz} = {\bf xzy}, xxz = 
xzx, yyz = yzy, {\bf yzx} = {\bf yxz}, \\
& & & & zxx, zyy, zzz, {\bf zxy} = {\bf zyx} \\
&&&$\pm 4_z,\overline{2}_{xy},\overline{2}_{-xy}$& xyz = xzy, xxz = xzx, yyz = 
yzy, yzx = yxz, \\
& & & & zxx, zyy, zzz, {\bf zxy} = {\bf zyx} \\
g&mm2&$1,2_z,\overline{2}_{xy},\overline{2}_{-xy}$&$\pm 
4_z,\overline{2}_x,\overline{2}_y$& xxz = xzx = yyz = yzy, {\bf xzy} = {\bf xyz} 
= {\bf yzx} = {\bf yxz}, \\
& & & & zxx = zyy, zzz, {\bf zxy} = {\bf zyx} \\
h&m&$1,\overline{2}_y$&$2_z,\overline{2}_x$& {\bf xxx}, {\bf xyy}, {\bf 
xzz}, xxz = xzx, yyz = yzy, \\
& & & & {\bf yyx} = {\bf yxy}, zxx, zzz, {\bf zzx} = {\bf zxz} \\
&&&$4_z,\overline{2}_{xy}$& xxx, xyy, xzz, xxz = xzx, yyz = yzy, \\
& & & & yyx = yxy, zxx, zzz, zzx = zxz \\
i&1&1&$2_z$& All the elements are allowed: \\
&&&& {\bf xxx}, {\bf xyy}, {\bf xzz}, xyz = xzy, xzx = xxz,\\
&&&& {\bf xxy} = {\bf xyx}, {\bf yxx}, {\bf yyy}, {\bf yzz}, yyz = yzy, \\
&&&& yzx = yxz, {\bf yxy} = {\bf yyx}, zxx, zyy, zzz, \\
&&&& {\bf zyz} = {\bf zzy}, {\bf zzx} = {\bf zxz}, zxy = zyx\\
&&&$\overline{2}_x$&{\bf xxx}, {\bf xyy}, {\bf xzz}, {\bf xyz} = {\bf xzy}, xzx 
= xxz,\\
&&&&xxy = xyx, yxx, yyy, yzz, yyz = yzy, \\
&&&&{\bf yzx} = {\bf yxz}, {\bf yxy} = {\bf yyx}, zxx, zyy, zzz, \\
&&&&zyz = zzy, {\bf zzx} = {\bf zxz}, {\bf zxy} = {\bf zyx}\\
&&&$\pm 4_z,\overline{2}_{xy},\overline{2}_{-xy}$& no information about the 
parity\\
j&m&$1,\overline{2}_{-xy}$&$2_z,\overline{2}_{xy}$& {\bf xxx} = {\bf yyy}, {\bf 
xyy} = {\bf yxx}, {\bf xzz} = {\bf yzz}, \\
& & & & xyz = yxz = xzy = yzx, xxz = xzx = yyz = yzy, \\
& & & & {\bf xxy} = {\bf yyx} = {\bf xyx} = {\bf yxy}, zxx = zyy, zzz, \\
& & & & {\bf zxz} = {\bf zzx} = {\bf zyz} = {\bf zzy}, zxy = zyx \\ 
&&&$4_z,\overline{2}_y$& {\em xxx} ={\pmb {\em yyy}}, {\em xyy} ={\pmb {\em 
yxx}}, {\em xzz} = {\pmb {\em yzz}}, \\
& & & & {\bf xyz} = {\bf yxz} = {\bf xzy} = {\bf yzx}, xxz = xzx = yyz = yzy, \\
& & & & {\pmb {\em xxy}} = {\pmb {\em xyx}} = {\em yyx} = {\em yxy}, zxx = zyy, 
zzz, \\
& & & & {\em zxz} = {\em zzx} = {\pmb {\em zyz}} = {\pmb {\em zzy}}, {\bf zxy} = 
{\bf zyx} \\ 
k&mm2&$1,2_z,\overline{2}_x,\overline{2}_y$&-& xxz = xzx, yyz = yzy, zxx, 
zyy, zzz \\
l&m&$1,\overline{2}_x$&$2_z,\overline{2}_y$& xzx = xxz, {\bf xxy} = {\bf 
xyx}, {\bf yxx}, {\bf yyy}, {\bf yzz}, \\
& & & & yyz = yzy, zxx, zyy, zzz, {\bf zyz} = {\bf zzy} \\
m&1&1&$2_z$& All the elements are allowed: \\
&&&& {\bf xxx}, {\bf xyy}, {\bf xzz}, xyz = xzy, xzx = xxz, {\bf xxy} = {\bf 
xyx}, \\
&&&& {\bf yxx}, {\bf yyy}, {\bf yzz}, yyz = yzy, yzx = yxz, {\bf yxy} = {\bf 
yyx}, \\
&&&& zxx, zyy, zzz, {\bf zyz} = {\bf zzy}, {\bf zzx} = {\bf zxz}, zxy = zyx\\
&&&$\overline{2}_x$&{\bf xxx}, {\bf xyy}, {\bf xzz}, {\bf xyz} = {\bf xzy}, xzx 
= xxz, xxy = xyx, \\
&&&&yxx, yyy, yzz, yyz = yzy, {\bf yzx} = {\bf yxz}, {\bf yxy} = {\bf yyx}, \\
&&&&zxx, zyy, zzz, zyz = zzy, {\bf zzx} = {\bf zxz}, {\bf zxy} = {\bf zyx}\\
n&2&$1,2_z$&$\overline{2}_x,\overline{2}_y$& {\bf xyz} = {\bf xzy}, xxz = 
xzx, yyz = yzy, {\bf yzx} = {\bf yxz}, \\
& & & &zxx, zyy, zzz, {\bf zxy} = {\bf zyx} \\
o&m&$1,\overline{2}_y$&$2_z,\overline{2}_x$&{\bf xxx}, {\bf xyy}, {\bf 
xzz}, xxz = xzx, yyz = yzy, \\
&&&&{\bf yyx} = {\bf yxy}, zxx, zyy, zzz, {\bf zzx} = {\bf zxz}\\
p&6mm&$1,2_z,\pm3_z, \pm6_z,6(\overline{2}_{\perp})$&-& xxz = xzx = yyz = 
yzy, zxx = zyy,  zzz \\
q&6&$1,2_z,\pm3_z,\pm6_z$&$\overline{2}_x,\overline{2}_y$&{\bf xyz} = {\bf 
xzy} = -{\bf yxz} = -{\bf yzx}, xxz = xzx = yyz = yzy,\\
&&&& zxx = zyy, zzz\\
r&3m&$1,\pm3_z,\overline{2}_y,\overline{2}_{S(xy)},\overline{2}_{S(-xy)}$
&-&zxx = zyy, xxz = xzx = yyz = yzy, zzz, \\
& & & & xxx = -xyy = -yxy = -yyx \\
s&1&1&$\overline{2}_y$& All the elements are allowed: \\
&&&& xxx, xyy, xzz, {\bf xyz} = {\bf xzy}, xzx = xxz, {\bf xxy} = {\bf xyx}, \\
&&&& {\bf yxx}, {\bf yyy}, {\bf yzz}, yyz = yzy, {\bf yzx} = {\bf yxz}, yxy =  
yyx, \\
&&&& zxx, zyy, zzz, {\bf zyz} = {\bf zzy}, zzx = zxz, {\bf zxy} = {\bf zyx}\\
t&m&$1,\overline{2}_y$&-& xxx, xyy, xzz, xxz = xzx, yyz = yzy, \\
& & & & yyx = yxy, zxx, zyy, zzz, zzx = zxz \\
u&3&$1,\pm3_z$&$\overline{2}_y$& xxx = -xyy = -yxy = -yyx, {\bf xyz} = {\bf 
xzy} = -{\bf yxz} = -{\bf yzx},\\
&&&& xzx = xxz = yyz = yzy, {\bf xxy} = {\bf xyx} = {\bf yxx} = -{\bf yyy},\\
&&&& zxx = zyy, zzz\\
w&1&1&-& All the elements are allowed \\
\end{tabular}
\end{table}

\begin{multicols}{2}
\narrowtext
\begin{table}
\caption[tab001A]{\label{tab1}SHG response for all spin configurations of the 
(001) surface of a fcc lattice~\cite{praca}. For the detailed description of the 
response types see Tab. \ref{keytab}. The configurations are depicted in Fig. 
\ref{fig1}.}
\begin{tabular}{cc}

configuration&key (response type)\\ \hline
& \\
para&a\\
ferro1&b\\
ferro2&c\\
ferro4&d\\
AF:& \\
a), b), e), o)&e\\
c), f)&f\\
i), k), m), p)&g\\
r)&a\\
\end{tabular}
\end{table}

\begin{table}
\caption[tab002A]{\label{t002}SHG response for all spin configurations of the 
(001) surface of a fcc lattice, with the spin structure of the second layer 
taken into account. For the detailed description of the response types see Tab. 
\ref{keytab}. For the confs. see Fig. \ref{fig1}.} 
\begin{tabular}{cc}
configuration&key (response type)\\ \hline
 & \\ 
para&a\\
ferro1&b\\
ferro2&c\\
ferro4&d\\
AF:&\\
ax), ox)&h\\
ay), oy), r)&e\\
bx), by), ex), ey)&b\\
c), fx), fy)&i\\
i)&j\\ 
k)&f\\
m), p)&c\\ 
\end{tabular}
\end{table}
%
%
\begin{table}
\caption[tab110A]{\label{tab2}SHG response for all spin configurations of the 
(110) surface of a fcc lattice \cite{praca}. For the detailed description of the 
response types see Tab. \ref{keytab}. The configurations are depicted in Fig. 
\ref{fig2}.} 
\begin{tabular}{cc}
configuration&key (response type)\\ \hline
 & \\
para&k\\
ferro1&l\\
ferro2&m\\
ferro3&n\\
ferro4&o\\
AF: & \\
a), b), c), g) - l)&k\\
d), e), f)&n\\
\end{tabular}
\end{table}   

\begin{table}
\caption[tab1111A]{\label{tab3}SHG response for all spin configurations of the 
(111) surface of a fcc lattice \cite{praca}. Only one monolayer is taken into 
account. For the detailed description of the response types see Tab. 
\ref{keytab}. The configurations are depicted in Fig. \ref{fig3}} 
\begin{tabular}{cc}
configuration&key (response type)\\ \hline
 & \\
para&p\\
ferro1&l\\
ferro3&o\\
ferro5&q\\
AF:& \\
a), i), k)&k\\
c), f)&n\\
\end{tabular}
\end{table}   

\begin{table}
\caption[tab1112A]{\label{tab4}SHG response for all spin configurations of the 
(111) surface of a fcc lattice \cite{praca}. More monolayers are taken into 
account. For the detailed description of the response types see Tab. 
\ref{keytab}. The configurations are depicted in Fig. \ref{fig3}.} 
\begin{tabular}{cc}
configuration&key (response type)\\ \hline
 & \\
para&r\\
ferro1&s\\
ferro3&t\\
ferro5&u\\
AF:& \\
a), i), k)&t\\
c), f)&u\\
\end{tabular}
\end{table}   

\begin{table}
\caption[tabB]{\label{tdis100}SHG response for all spin configurations of the 
(001) surface of a fcc lattice, distorted to a rhombohedral structure. For the 
detailed description of the response types see Tab. \ref{keytab}. For the 
surface structure see Fig. \ref{fdis111}, for the spin configurations see Fig. 
\ref{fig1}.} 
\begin{tabular}{cc}
configuration&key (response type)\\ \hline
 &\\ 
para&k\\
ferro1&m\\
ferro2&o\\
ferro3&l\\
ferro4&n\\
AF:&\\
a), b) - h), o)&n\\
i) - n), p) - r)&k\\
\end{tabular}
\end{table}   

\begin{table}
\caption[tab1111B]{\label{tdis1111}SHG response for all spin configurations of 
the (111) surface of a fcc lattice, distorted to a rhombohedral structure. Only 
one monolayer is taken into account. For the detailed description of the 
response types see Tab. \ref{keytab}. For the surface structure see Fig. 
\ref{fdis111}, for the spin configurations see Fig. \ref{fig3}.} 
\begin{tabular}{cc}
configuration&key (response type)\\ \hline
 & \\ 
para&k\\
ferro1, ferro4&l\\
ferro2&m\\
ferro3&o\\
ferro5&n\\
AF: & \\
a), k)&k\\
b) - j), l), m)&n\\
\end{tabular}
\end{table}   

\begin{table}
\caption[tab1112B]{\label{tdis1112}SHG response for all spin configurations of 
the (111) surface of a fcc lattice, distorted to a rhombohedral structure. More 
monolayers are taken into account. For the detailed description of of the 
response types see Tab. \ref{keytab}. For the surface structure see Fig. 
\ref{fdis111}, for the spin configurations see Fig. \ref{fig3}.} 
\begin{tabular}{cc}
configuration&key (response type)\\ \hline
 & \\ 
para&t\\
ferro1, ferro2, ferro4, ferro5&s\\
ferro3&t\\
AF: & \\
a), i), k)&s\\
b) - h), j), l), m) &t\\
\end{tabular}
\end{table}   

\begin{table}
\caption[tab001C]{\label{tnierow100}SHG response for all spin configurations of 
the (001) surface of a fcc lattice, with one atom distinguished. 
For the detailed description of the response types see Tab. \ref{keytab}. For 
the surface arrangement see Fig. \ref{fnierow}. For the confs. see Fig. 
\ref{fig1}.}
\begin{tabular}{cc}
configuration&key (response type)\\ \hline
 & \\ 
para&a\\
ferro1&b\\
ferro2&c\\ 
ferro4&d\\
AF:& \\
a), o)&h\\
b), e)&b\\
c)&f\\
f)&i\\
i), m), p)&e\\
k)&j\\ 
r)&d\\
\end{tabular}
\end{table}

\begin{table}
\caption[tab110C]{\label{tnierow110}SHG response for all spin configurations of 
the (110) surface of a fcc lattice, with one atom distinguished. For detailed 
description of response types see Tab. \ref{keytab}. For the surface arrangement 
see Fig. \ref{fnierow}. For the confs. see Fig. 
\ref{fig2}.}
\begin{tabular}{cc}
configuration&key (response type)\\ \hline
 & \\ 
para&k\\
ferro1&l\\
ferro2&m\\
ferro3&n\\
ferro4&o\\
AF: & \\
a)&l\\
b), c), h), i), k), l)&k\\
d)&1m\\
e), f), g)&n\\
j)&o\\
\end{tabular}
\end{table}   

\begin{table}
\caption[tab1111C]{\label{tnierow1111}SHG response for all spin configurations 
of the (111) surface of a fcc lattice, with one atom distinguished. Only one 
monolayer taken into account. For the detailed description of the response types 
see Tab. \ref{keytab}. For the surface arrangement see Fig. \ref{fnierow}. For 
the confs. see Fig. \ref{fig3}.}
\begin{tabular}{cc}
configuration&key (response type)\\ \hline
 & \\
para&p\\
ferro1&l\\
ferro3&o\\
ferro5&q\\
AF:& \\
a)&l\\
c), f)&m\\
i)&o\\
k)&n\\
\end{tabular}
\end{table}   

\begin{table}
\caption[tab1112C]{\label{tnierow1112}SHG response for all spin configurations 
of the (111) surface of a fcc lattice, with one atom distinguished. More 
monolayers are taken into account. For the detailed description of the response 
types see Tab. \ref{keytab}. For the surface arrangement 
see Fig. \ref{fnierow}. 
For the confs. see Fig. \ref{fig3}.}
\begin{tabular}{cc}
configuration&key (response type)\\ \hline
 & \\
para&3r\\
ferro1&s\\
ferro3&t\\
ferro5&u\\
AF:& \\
a), c), f), k)&s\\
i)&t\\
\end{tabular}
\end{table}   

\begin{table}
\caption[tab001D]{\label{ttlen100}SHG response for all spin configurations of 
the (001) surface of a fcc lattice, with a distortion of oxygen sublattice. For 
the detailed description of the response types see Tab. \ref{keytab}. For the 
surface arrangement see Fig. \ref{ftlen}. For the confs. see Fig. \ref{fig1}.} 
\begin{tabular}{cc}
configuration&key (response type)\\ \hline
 & \\ 
para&a\\
ferro1&b\\
ferro2&c\\ 
ferro4&d\\
AF:& \\
a), o)&h\\
b), e)&b\\
c), f)&i\\ 
i), k)&c\\ 
m)&j\\ 
p)&e\\
r)&d\\
\end{tabular}
\end{table}   

\begin{table}
\caption[tab110D]{\label{ttlen110}SHG response for all spin configurations of 
the (110) surface of a fcc lattice, with oxygen sublattice distorted. For the 
detailed description of the response types see Tab. \ref{keytab}. For the 
surface arrangement see Fig. \ref{ftlen}. For the confs. see Fig. \ref{fig2}.} 
\begin{tabular}{cc}
configuration&key (response type)\\ \hline
 & \\ 
para&k\\
ferro1&l\\
ferro2&m\\
ferro3&n\\
ferro4&o\\
AF: & \\
a), b), g), h), k), l)&k\\
c)&o\\
d), e), i), j)&n\\
f)&m\\
\end{tabular}
\end{table}   

\begin{table}
\caption[tab111D]{\label{ttlen111}SHG response for all spin configurations of 
the (111) surface of a fcc lattice, with oxygen sublattice distorted. For the 
detailed description of the response types see Tab. \ref{keytab}. For the 
surface arrangement see Fig. 
\ref{ftlen}. For the confs. see Fig. \ref{fig3}.}
\begin{tabular}{cc}
configuration&key (response type)\\ \hline
 & \\ 
para&u\\
ferro1, ferro3&w\\
ferro5&u\\
AF:&\\
All confs.&w\\
\end{tabular}
\end{table}   
\end{multicols}

\begin{thebibliography} {99}
\bibitem{RemTanabe} The authors perform the analysis of $\chi^{(2\omega)}$ in 
the frequency space. There are no reasons to neglect the damping in this 
approach. On the contrary, if the dissipation is neglected, then, for instance, 
the Kramers-Kronig relations cannot be applied. In real time, on the 
other hand, the dissipation does not take place, as stated in \cite {Bassani}.
\bibitem{par111} The notion of the parity operation requires that the original 
situation is restored after applying the operation at most twice. Obviously, 
$2_z$ is such an operation, and for certain configurations $4_z$ satisfies this 
criterion as well (where $2_z$ is just a symmetry operation). However, neither 
$3_z$ nor $6_z$ have this property - they must be applied at least 3 times to 
restore the original situation. Concerning the tensor elements, for $2_z$ and 
mirror-operations a given tensor element may change its sign or not. Upon the 
application of the $4_z$ operation it may become another tensor element (i.e. 
its x indices become y and vice versa) while its sign can change or remain 
unchanged. Upon the $3_z$ operation, however, a single tensor element gets 
``split'' into several tensor elements, and the notion of the conservation (or 
not) of the sign of the tensor element looses its sense. The tensor elements are 
described in the cartesian coordinate system, where quarters are the elementary 
entities. Rotation by an angle other than $90^\circ$ and its multiples cannot be 
described as interchanging the axes and a possible modification of their signs. 
However, it is possible to treat the parity of tensor elements in the $3_z$ and 
$6_z$ operation in a more complicated way.
\bibitem{leftout} If a spin structure is not described within this subsection 
(nor in the tables relevant to this subsection), it is a domain of the last 
displayed configuration that precedes the omitted one. This applies to all the 
subsections.
\bibitem{4atoms} Although our primitive cell contains 4 magnetic atoms they are 
not placed on one line, and thus we still have a two-sublattice antiferromagnet.
\bibitem{domeny} It is also possible for different bulk domains to yield the 
same spin ordering at the surface.
\end{thebibliography}
\end{document}